\documentclass[aps,10pt,prb,bibliography,citeautoscript,superscriptaddress,twocolumn]{revtex4-2}
\usepackage[T1]{fontenc}
\usepackage{graphicx}
\usepackage{dcolumn}
\usepackage{amsmath,amssymb,amsfonts,bm,dsfont,units}
\usepackage{hyperref}
\hypersetup{colorlinks=true,citecolor=blue,linkcolor=blue,urlcolor=blue,pdfstartview=FitH}
\usepackage{braket}
\usepackage{float,latexsym}
\usepackage{multirow}
\usepackage[normalem]{ulem} % for \sout
\usepackage[caption=false]{subfig} % for subfloat

\usepackage{array}
\usepackage{xcolor}
\usepackage{graphicx} 
\usepackage{hhline,booktabs}
\usepackage{makecell}

\newcommand{\sfigure}[2]{Figure~\hyperref[#1]{\ref{#1}(#2)}}
\newcommand{\sfigref}[2]{Fig.~\hyperref[#1]{\ref{#1}(#2)}}

\usepackage{color}
\definecolor{dkgreen}{rgb}{0,0.5,0}
\definecolor{midnightblue}{rgb}{0.39,0.58,0.93}

\definecolor{kspink}{RGB}{200,0,200}
\definecolor{appendixgreen}{RGB}{9, 145, 58}

\newcommand{\comment}[1]{}{}
\begin{document}

\title{Assembling Kitaev honeycomb spin liquids from arrays of 1D symmetry protected topological phases}

\author{Yue Liu}
  \affiliation{Department of Physics, California Institute of Technology, Pasadena, CA 91125, USA}
  \affiliation{Institute for Quantum Information and Matter, California Institute of Technology, Pasadena, CA 91125, USA}
 \author{Nathanan Tantivasadakarn}
  \affiliation{Walter Burke Institute for Theoretical Physics, California Institute of Technology, Pasadena, CA 91125, USA}
  \affiliation{Department of Physics, California Institute of Technology, Pasadena, CA 91125, USA}
\author{Kevin Slagle}
  \affiliation{Department of Physics, California Institute of Technology, Pasadena, CA 91125, USA}
  \affiliation{Institute for Quantum Information and Matter, California Institute of Technology, Pasadena, CA 91125, USA}
  \affiliation{Walter Burke Institute for Theoretical Physics, California Institute of Technology, Pasadena, CA 91125, USA}
 \affiliation{Department of Electrical and Computer Engineering,
Rice University, Houston, Texas 77005 USA}
 \author{David F. Mross}
\affiliation{Department of Condensed Matter Physics, Weizmann Institute of Science, Rehovot 7610001, Israel}
     \author{Jason Alicea}
  \affiliation{Department of Physics, California Institute of Technology, Pasadena, CA 91125, USA}
  \affiliation{Institute for Quantum Information and Matter, California Institute of Technology, Pasadena, CA 91125, USA}
 \affiliation{Walter Burke Institute for Theoretical Physics, California Institute of Technology, Pasadena, CA 91125, USA}

\date{\today}

	\begin{abstract}
		The Kitaev honeycomb model, which is exactly solvable by virtue of an extensive number of conserved quantities, supports a gapless quantum spin liquid phase as well as gapped descendants relevant for fault-tolerant quantum computation. We show that the anomalous edge modes of 1D cluster-state-like symmetry protected topological (SPT) phases provide natural building blocks for a variant of the Kitaev model that enjoys only a \emph{subextensive} number of conserved quantities. The symmetry of our variant allows a single additional nearest-neighbor perturbation, corresponding to an anisotropic version of the $\Gamma$ term studied in the context of Kitaev materials. We determine the phase diagram of the model using exact diagonalization. Additionally, we use DMRG to show that the underlying 1D SPT building blocks can emerge from a ladder Hamiltonian exhibiting only two-spin interactions supplemented by a Zeeman field. Our approach may inform a new pathway toward realizing Kitaev honeycomb spin liquids in spin-orbit-coupled Mott insulators.
	\end{abstract}
	\maketitle

\section{Introduction}

Exactly solvable models play an essential part in the understanding of many strongly interacting, fractionalized phases of matter (see, e.g., Refs.~\onlinecite{KitaevAnyons,XieCohomology,Levin2005,Walker2011,VijayFracton,fractonReview}).  Typically, exact solvability descends from an extensive set of conserved quantities---i.e., whose number scales with system size---exhibited by a microscopic Hamiltonian.  Such extensive conserved quantities can simultaneously quell competition from conventional orders while enabling a minimalist description of exotic ground states and the nontrivial emergent excitations that they host.  Experimental platforms, by contrast, generically exhibit vastly fewer conserved quantities associated with a set of physical global symmetries that is independent of system size.  Nevertheless, exactly solvable models can inform searches for materials governed by Hamiltonians that are sufficiently `nearby' to realize the same universal properties.

As an important example, the exactly solvable Kitaev honeycomb model \cite{KitaevAnyons} captures a family of quantum spin liquid phases for spin-1/2 degrees of freedom arranged on a honeycomb lattice.  Here, special bond-dependent spin interactions are incorporated such that the Hamiltonian preserves local, mutually commuting multi-spin operators associated with each hexagonal plaquette.  The virtue of these conserved quantities manifests upon employing a Majorana representation of the spins: the Hamiltonian then maps to Majorana fermions coupled to a $\mathbb{Z}_2$ gauge field whose flux, crucially, has \emph{no dynamics}.   In any fixed flux sector, the Hamiltonian moreover reduces to a free fermion problem, whose wavefunctions and energies can be efficiently solved.
The exact solution reveals a gapless $\mathbb{Z}_2$ spin liquid phase hosting emergent massless Dirac fermions born from a purely bosonic spin system.  Additionally, the model supports toric code and non-Abelian spin liquids---both of which are sought for fault-tolerant quantum computation---as proximate descendants that arise upon gapping the emergent fermions.  

Seminal work by Jackeli and Khaliullin identified a promising route to material realizations \cite{Jackeli2009}. 
Specifically, they predicted that a family of spin-orbit-coupled Mott insulators---now dubbed `Kitaev materials'---exhibits precisely the bond-dependent spin interactions from the Kitaev honeycomb model; however, perturbations inevitably exist that spoil exact solvability, endow $\mathbb{Z}_2$ fluxes with dynamics, and promote competing orders.  Indeed, at zero magnetic field most Kitaev materials magnetically order at low temperatures \cite{Savary2016,Trebst2017,Winter2017,Hermanns2018,Janssen2019,Takagi2019,Motome2020}, indicating that such perturbations are sufficiently severe to destabilize the gapless quantum spin liquid.  Signatures of fractionalization have, nevertheless, been reported \cite{Wang2017,Banerjee2017,Kasahara2018,Kasahara2018nat,Takagi2019,Wang2020,Yokoi2021,Bruin2022}, though the experimental situation remains unsettled \cite{Bachus2020,Yamashita2020,Chern2021,Bachus2021,Czajka2021,JCCM2021}.  

Devising alternative spin-anisotropic microscopic Hamiltonians that capture similar phenomenology to the Kitaev honeycomb model could potentially expand the landscape of candidate spin liquid materials \cite{thomson2018,Sagi2019,Verresen22,sahay2023,Verresen23,Grushin2023}.  To this end, we propose consideration of models that lie intermediate between exactly solvable and generic, experimentally realistic Hamiltonians, in the sense of enjoying a number of conserved quantities that grows with system size, but only \emph{subextensively}.  It is natural to anticipate that subextensive conserved quantities, while insufficient for exact solvability, can still more efficiently suppress competing orders relative to generic Hamiltonians---thus providing fertile ground for spin-liquid explorations.  

\begin{figure}[h]
    \centering
    \includegraphics[width=\linewidth]{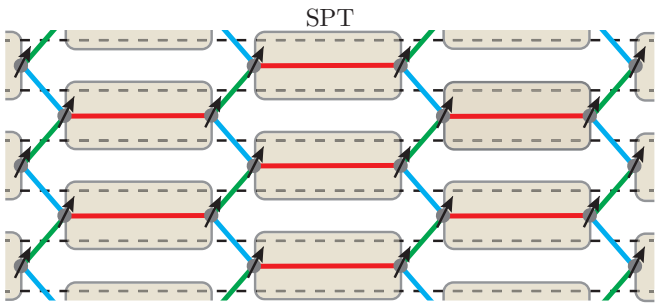}
    \caption{Kitaev honeycomb model variant built from an array of symmetry protected topological phases (SPT's) that correspond to reflection-symmetric cousins of the one-dimensional cluster state.  Each SPT building block hosts anomalous spin-1/2 edge states, and arises microscopically from a spin-1/2 ladder model that enjoys separate spin-flip symmetries for each leg---collectively yielding a subextensive number of conserved quantities for the array.      Symmetry-allowed interactions among anomalous edge spins connected by green, blue, and red bonds generate the Kitaev honeycomb model perturbed by a single anisotropic spin-exchange term.}
    \label{fig:array}
\end{figure}

We substantiate the expectation above by introducing a microscopic spin-liquid model with subextensive conserved quantities associated with flipping all spins in any given `row' of the lattice.  Figure~\ref{fig:array} sketches our construction, which is based on an array of one-dimensional symmetry protected topological phases (SPTs).  Each SPT building block corresponds to a reflection-symmetric counterpart of the `cluster state'---originally introduced in the context of measurement-based quantum computation \cite{Raussendorf01,Verresen17}---and hosts anomalous spin-1/2 edge states.  In our setup the SPT arises microscopically from a spin-1/2 ladder Hamiltonian involving relatively simple one- and two-spin interactions.  When arrayed as in Fig.~\ref{fig:array}, the anomalous edge spins hybridize along a honeycomb lattice as illustrated by the green, blue, and red bonds.  Remarkably, the subextensive conserved quantities built into our Hamiltonian constrain the interactions along these bonds to exactly the structure of couplings in the Kitaev honeycomb model---up to a \emph{single} perturbation corresponding to a spatially anisotropic version of the $\Gamma$ term present for Kitaev materials \cite{Rau2014}.  This single perturbation mediates restricted dynamics for $\mathbb{Z}_2$ fluxes in which they can propagate along only one direction of the lattice.  The gapless $\mathbb{Z}_2$ spin liquid generically survives a finite threshold of these nontrivial flux processes, and we show explicitly that our setup can reside below that threshold.

Although our construction reduces at low energies to a variant of the Kitaev honeycomb model, we stress that the underlying microscopic interactions are entirely different.  This observation raises the hope that similarly utilizing subextensive conserved quantities may in the future unearth new, realistic Hamiltonian targets for emulation in experiments.  Our work also highlights nearer-term opportunities for experimentally realizing individual SPT building blocks, given the rather simple structure of the required Hamiltonian.

We organize the remainder of the paper as follows.  Section~\ref{ClusterStateSec} reviews the canonical one-dimensional cluster state and introduces our reflection-symmetric extension in the context of effective spin-1 models.   Section~\ref{sec:realistic_implementation} then identifies a parent Hamiltonian for the reflection-symmetric cluster state in a more realistic spin-1/2 ladder setup.  In Sec.~\ref{KitaevVariant} we examine the SPT array from Fig.~\ref{fig:array} and establish the conditions for realizing the gapless quantum spin liquid from the Kitaev honeycomb model.  A summary and discussion of our main findings is given in Sec.~\ref{Discussion}.  Numerous appendices detail supplementary results and background information.

\section{Reflection-symmetric 1D cluster state SPT}
\label{ClusterStateSec}

\subsection{Canonical cluster state review}
\label{ClusterStateReview}

\begin{figure}[h]
    \centering
    \includegraphics[width=0.95\linewidth]{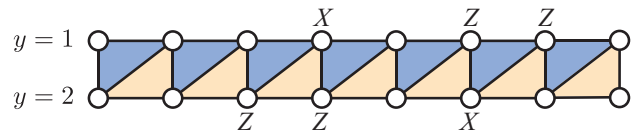}
    \caption{Schematic of the canonical cluster state SPT on a square ladder.  The model contains $ZXZ$ terms along the two types of shaded triangles---which do not transform into one another under inter-chain reflections.  }
    \label{fig:twochain}
\end{figure}

We begin by considering spin-1/2 degrees of freedom on a square ladder (Fig.~\ref{fig:twochain}), governed by the Hamiltonian 
\begin{equation}
    H_{\rm cluster} = -J\sum_{j=1}^{N-1}(Z_{j,1} X_{j,2} Z_{j+1,1} + Z_{j,2} X_{j+1,1} Z_{j+1,2}).
    \label{Hcluster}
\end{equation}
Here $N$ denotes the number of sites in each chain comprising the ladder, $X_{j,y}$ and $Z_{j,y}$ are Pauli operators acting on site $j$ in chain $y$, and we assume $J>0$.  
Equation~\eqref{Hcluster}  
famously hosts a $\mathbb{Z}_2 \times \mathbb{Z}_2$ cluster state SPT phase \cite{Raussendorf01,Verresen17}.  In particular, the two $\mathbb{Z}_2$ symmetries are generated by 
\begin{equation}
  G_1 = \prod_j X_{j, 1},~~~~G_2 = \prod_j X_{j, 2},    
  \label{Z2s}
\end{equation} 
and correspond to invariance of $H_{\rm cluster}$ under globally flipping the spins in either chain $y = 1$ or 2.

All terms in $H_{\rm cluster}$ commute with each other and square to the identity; hence ground states must exhibit 
\begin{equation}
  Z_{j,1} X_{j,2} Z_{j+1,1} = Z_{j,2} X_{j+1,1} Z_{j+1,2} = +1     
  \label{GroundStateCondition}
\end{equation} 
for all $j$.
These ground-state conditions can be combined to define string order parameters characterizing the cluster state SPT:
\begin{equation}
    Z_{j, 1} \left( \prod_{k=j}^{j'-1} X_{k, 2} \right)Z_{j', 1} = \prod_{k = j}^{j'-1} Z_{k,1} X_{k,2} Z_{k+1,1} =1
    \label{eq:stringop1}
\end{equation}
for arbitrary $j'>j$, and similarly 
\begin{equation}
    Z_{j, 2} \left( \prod_{k=j+1}^{j'} X_{k, 1}\right)Z_{j', 2} =1.
    \label{eq:stringop2}
\end{equation}
Moreover, the operators
\begin{equation}
    \mathcal{X}_L = Z_{1,1},~~~~ \mathcal{Y}_L = X_{1,1} Z_{1,2}
\end{equation}    
localized at the left edge and 
\begin{equation} 
    \mathcal{X}_R = Z_{N,2},~~~~ \mathcal{Y}_R = X_{N,2} Z_{N,1}
\end{equation}
localized at the right edge preserve the ground-state conditions in Eq.~\eqref{GroundStateCondition}.  These operators define the anomalous edge zero modes characteristic of the cluster-state SPT and together span a four-fold ground state degeneracy.  Modulo corrections that decay exponentially with system size, this degeneracy is immune to arbitrary local perturbations that preserve the two $\mathbb{Z}_2$ symmetries protecting the SPT order.  (Note that in our definitions above $\mathcal{X}_L$ and $\mathcal{Y}_R$ are odd under $G_1$, whereas $\mathcal{Y}_L$ and $\mathcal{X}_R$ are odd under $G_2$.  Throughout we adopt similar conventions since they facilitate connection to the Kitaev honeycomb model in Sec.~\ref{KitaevVariant}.)

\subsection{Reflection-symmetric extension}
\label{sec:spin1}

The canonical cluster state Hamiltonian from Eq.~\eqref{Hcluster} lacks inter-chain reflection symmetry since the two sets of  triangles on which the $ZXZ$ terms act do not transform into one another upon swapping the two chains of the square ladder; see Fig.~\ref{fig:twochain}.
Here we wish to introduce a cluster state variant that preserves inter-chain reflections---yielding an additional symmetry denoted $\mathbb{Z}^{\rm ref}_2$.  Including the $\mathbb{Z}_2$ symmetries associated with global spin flips in each chain, the symmetry group then extends to the dihedral group $D_4 = (\mathbb{Z}_2 \times \mathbb{Z}_2) \rtimes\mathbb{Z}^{\rm ref}_2$. The classification of 1D SPT phases protected by such a group is given by $H^2(D_4,U(1))=\mathbb Z_2$. Moreover, the nontrivial SPT here persists upon restricting the symmetry to the $\mathbb Z_2 \times \mathbb Z_2$ subgroup, and is in the same phase as the cluster state. In principle, a commuting-projector model for which the cluster state SPT phase enjoys an additional inter-chain reflection symmetry should therefore exist. 

We instead follow a more illuminating route to the reflection-symmetric cluster state that directly connects with the realistic spin Hamiltonians that we simulate in Sec.~\ref{sec:realistic_implementation}.  As an initial step we reduce the spin-1/2 ladder to a spin-1 chain by introducing a reflection-symmetric parent Hamiltonian whose dominant term is
\begin{equation}
    H_{\rm triplet}  = -\lambda \sum_{j} (X_{j,1}+X_{j,2}-X_{j,1}X_{j,2})
    \label{H_triplet}
\end{equation}
with $\lambda>0$.  For the pair of sites on a rung labeled by $j$, let us denote the $X$-basis eigenstates by $\ket{X_{j,1} = 1,X_{j,2}= 1} = \ket{\rightarrow,\rightarrow}$, $\ket{X_{j,1} = 1,X_{j,2} = -1}=\ket{\rightarrow,\leftarrow}$, etc.  Equation~\eqref{H_triplet} penalizes the $\ket{\leftarrow,\leftarrow}$ configuration but leaves
a triplet of degenerate ground states that we group as
\begin{align}
  \ket{1} &=  \frac{\ket{\leftarrow,\rightarrow} - i \ket{\rightarrow,\leftarrow}}{\sqrt{2}} 
  \\
  \ket{0} &= \ket{\rightarrow,\rightarrow}
  \\
  \ket{-1} &= \frac{\ket{\leftarrow,\rightarrow} + i \ket{\rightarrow,\leftarrow}}{\sqrt{2}} .
\end{align}
We project onto the latter manifold and describe the resulting low-energy sector with spin-1 operators defined by
\begin{align}
    S^x_j &= \frac{1}{\sqrt{2}} \left(
			\begin{smallmatrix}
				0 & 1 & 0\\
				1 & 0 & 1 \\
                0 & 1 & 0\\
			\end{smallmatrix}\right), 
   S^y_j = \frac{i}{\sqrt{2}} \left(
			\begin{smallmatrix}
				0 & -1 & 0\\
				1 & 0 & -1 \\
                0 & 1 & 0\\
			\end{smallmatrix}\right),
   S^z_j = \left(
			\begin{smallmatrix}
				1 & 0 & 0\\
				0 & 0 & 0 \\
                0 & 0 & -1\\
			\end{smallmatrix}\right)
\end{align}
under the basis $(\ket{1},\ket{0},\ket{-1})$.  
In this spin-1 representation, the two $\mathbb{Z}_2$ symmetries from Eq.~\eqref{Z2s} map to global $\pi$ spin rotations about the $S^y$ and $S^x$ directions, i.e.,
\begin{equation}
  G_1 = e^{i \pi \sum_j S^y_j},~~~~G_2 = e^{i \pi \sum_j S^x_j},
  \label{G12}
\end{equation}
while inter-chain reflection symmetry corresponds to a $\pi$ rotation about the diagonal between the $S^x$ and $S^y$ directions,
\begin{equation}
    G_{\rm ref}  %\prod_j e^{i \frac{\pi}{2} S^z_j} e^{i \pi S^y_j} \textcolor{midnightblue}{ 
    = e^{i \pi \sum_j (S^x_j + S^y_j)/\sqrt{2}}.
    \label{Gref}
\end{equation}
[We dropped an unimportant factor of $e^{i\pi \sum_j}$ on the right side of Eqs.~\eqref{G12} through \eqref{Gref}.]  For later use, Table~\ref{tab.projection} lists the projection of various spin-1/2 ladder operators onto the effective spin-1 problem obtained above.  

\begin{table}[h]
\centering
\setlength{\extrarowheight}{2pt} 
\begin{tabular}{|c|c|}
\hline
 Spin-1/2 ladder operators & Projection onto spin-1 chain\\
 \hline
 $Z_{j,1}$ & $S^x_j$\\
 \hline
 $Z_{j,2}$ & $S^y_j$\\
 \hline
 $Y_{j,1} Z_{j,2}$ or $-Z_{j,1}Y_{j,2}$ & $S^z_j$\\
 \hline
 $X_{j,1}$ & $2(S^y_j)^2 - 1$\\
 \hline
 $X_{j,2}$ & $2(S^x_j)^2 - 1$\\
 \hline
 % $X_{j,1} + X_{j,2}$ & $2-2(S^z_j)^2$\\
 % \hline
 $X_{j,1} X_{j,2}$ & $1-2(S^z_j)^2$\\
 \hline
\end{tabular}
\caption{Projection of select spin-1/2 operators onto the effective spin-1 problem defined in Sec.~\ref{sec:spin1}.
}
\label{tab.projection}
\end{table}

At this point the conceptually simplest way to access a reflection-symmetric cluster state SPT is to put the effective spin-1 degrees of freedom into the ground state of the AKLT Hamiltonian
\begin{equation}
    H_{\rm AKLT} = \sum_{j} \left[\vec{S}_j \cdot \vec{S}_{j+1}+\frac{1}{3}\left(\vec{S}_j \cdot \vec{S}_{j+1}\right)^2\right],
    \label{H_AKLT}
\end{equation}
which clearly preserves both $G_{1,2}$ and inter-chain reflections $G_{\rm ref}$.  
(See Appendix~\ref{AKLTreview} for a brief review of the ground state and edge structure of $H_{\rm AKLT}$.)  The AKLT model realizes an SPT phase---aka Haldane phase---characterized by string order parameters
\begin{equation}
\mathcal{O}^{\alpha}_{\rm AKLT} = \left\langle S^\alpha_j \left[\prod_{k=j+1}^{j'-1}\exp{(i \pi S^\alpha_k)} \right] S^\alpha_{j'}\right \rangle \neq 0
\label{AKLTstring}
\end{equation}
for $\alpha = x,y,z$.
Like the canonical cluster state reviewed in Sec.~\ref{ClusterStateReview}, the AKLT SPT phase also hosts anomalous edge zero modes that span a four-fold ground state degeneracy on an open chain.
Reference~\onlinecite{Verresen17} previously established a link between the \emph{canonical} cluster state Hamiltonian [Eq.~\eqref{Hcluster}] and the AKLT chain.  Our construction, by contrast, incorporates an additional microscopic symmetry.  In particular, the edge zero mode operators now exhibit well-defined transformations under inter-chain reflection.  Equations~\eqref{G12} and \eqref{Gref} correspond to elements of continuous spin rotation symmetry built into $H_{\rm AKLT}$; consequently, one can back out the relevant edge zero mode symmetry properties using the fact that the edge spins transform projectively under global O$(3)$ transformations,
\begin{align}
e^{i \pi \sum_j \hat n \cdot \vec S_j}|\text{AKLT}\rangle =e^{i \frac{\pi}{2} \hat n \cdot \vec \sigma_\text{left}}e^{i \frac{\pi}{2} \hat n \cdot \vec \sigma_\text{right}}|\text{AKLT}\rangle,
\end{align}
where $\ket{\text{AKLT}}$ belongs to the ground state manifold and $\vec \sigma_{\rm left}$ and $\vec \sigma_{\rm right}$ denote Pauli matrices that act on the anomalous edge spin-1/2's.  Appendix~\ref{AKLTreview} details the analysis, the results of which are summarized in Table~\ref{tab.edgetransform}. (Note again that, for ease of connecting with the Kitaev honeycomb model later on, we adopt a convention where $\mathcal{X}_L$ and $\mathcal{Y}_R$ are odd under $G_1$ while $\mathcal{X}_R$ and $\mathcal{Y}_L$ are odd under $G_2$.)  

\begin{table}[h]
\centering
\setlength{\extrarowheight}{2pt}
\begin{tabular*}{\linewidth}{@{\extracolsep{\fill}}|c|c|c|c|c|c|c|}
\hline
 & $\mathcal{X}_L$~ & $\mathcal{Y}_L$~ & $\mathcal{Z}_L$~ & $\mathcal{X}_R$~ & $\mathcal{Y}_R$~ & $\mathcal{Z}_R$~\\
 \hline
 ~~~$G_1$~~~& $-\mathcal{X}_L$~ & $\mathcal{Y}_L$~ & $-\mathcal{Z}_L$~ & $\mathcal{X}_R$~ & $-\mathcal{Y}_R$~ & $-\mathcal{Z}_R$~ \\
 \hline
 ~~~$G_2$~~~ & $\mathcal{X}_L$~ & $-\mathcal{Y}_L$~ & $-\mathcal{Z}_L$~ & $-\mathcal{X}_R$~ & $\mathcal{Y}_R$~ & $-\mathcal{Z}_R$~ \\
 \hline
 ~~~$G_{\text{ref}}$~~~ & $\mathcal{Y}_L$~ & $\mathcal{X}_L$~ & $-\mathcal{Z}_L$~ & $\mathcal{Y}_R$~ & $\mathcal{X}_R$~ & $-\mathcal{Z}_R$~ \\
 \hline
\end{tabular*}
\caption{Symmetry transformations of the edge zero mode operators in the reflection-symmetric cluster state. 
}\label{tab.edgetransform}
\end{table}

The conceptual simplicity afforded by Eq.~\eqref{H_AKLT} comes with a drawback: In the original spin language, realizing $H_{\rm AKLT}$ requires somewhat baroque four-spin terms (see Table~\ref{tab.projection} and App.~\ref{app.akltfrom2chain}).  Thus we consider the alternative spin-1 model 
\begin{align}
    H_{\text{spin-1}} &= \sum_j [J(S^x_j S^x_{j+1} + S^y_j S^y_{j+1})
    \nonumber \\ 
    &+ J'\left(S^x_j S^x_{j+2} + S^y_j S^y_{j+2}\right) 
    + D (S^z_j)^2]
    \label{H_spin1}
\end{align}
that also preserves $G_{1,2}$ and $G_{\rm ref}$.   Equation~\eqref{H_spin1} features first- and second-neighbor easy-plane interactions 
that we take to be antiferromagnetic $(J,J'\geq 0)$, 
together with single-ion anisotropy that locally favors either the $S_j^z = 0$ state (for $D>0$) or the $S_j^z = \pm 1$ doublet (for $D<0$).  Notably, all of these Hamiltonian terms can arise microscopically from one- and two-spin interactions in the original spin-1/2 ladder; see again Table~\ref{tab.projection}.  References~\onlinecite{tonegawa1995effect,kaburagi1999spin} numerically studied the phase diagram of $H_{\text{spin-1}}$. When $J'=D=0$, a gapless Luttinger liquid emerges. Starting from this point, turning on arbitrarily weak $J'>0$ stabilizes the gapped SPT phase captured by the AKLT model, i.e., the reflection-symmetric cluster state in our context.  The SPT order persists until $J'/J \approx 0.48$, and moreover withstands a finite range of single-ion anisotropy $D$ by virtue of the gap.  Our own iDMRG simulations 
support the structure of the phase diagram identified in these early works; see Appendix~\ref{app.Hspin1} for details. 

\section{Simplified parent Hamiltonian of the reflection-symmetric cluster state}
\label{sec:realistic_implementation}

\subsection{Model and phase diagram}
\begin{figure}[h]
    \centering
    \includegraphics[width=\linewidth]{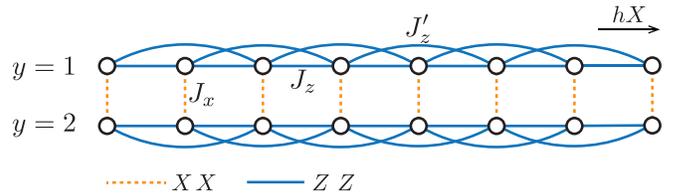}
    \caption{Schematic of the spin-1/2 ladder model in Eq.~\eqref{eq:twochain}, which preserves separate spin-flip symmetries for each of the two chains as well as inter-chain reflection symmetry.  }
    \label{fig:twochainH}
\end{figure}

Armed with the insights from Sec.~\ref{sec:spin1}, we now consider the spin-1/2 ladder model 
\begin{equation}
\begin{aligned}
    H = \sum_{y=1,2}\bigg{[}J_z&\sum_{j=1}^{N-1}Z_{j,y}Z_{j+1,y} + J_z'\sum_{j=1}^{N-2}Z_{j,y}Z_{j+2,y}\\
     - h &\sum_{j=1}^{N}X_{j,y}\bigg{]} + J_x\sum_{j=1}^{N} X_{j,1}X_{j,2}
\end{aligned}\label{eq:twochain}
\end{equation}
that encodes a transverse field $h$, antiferromagnetic intra-chain nearest-neighbor ($J_z>0$) and second-neighbor ($J_z'\geq 0$) $ZZ$-type Ising interactions, and antiferromagnetic inter-chain $XX$-type Ising interactions ($J_x\geq 0$); see Fig.~\ref{fig:twochainH}.  Equation~\eqref{eq:twochain} preserves inter-chain reflection symmetry and---due to the form of the inter-chain coupling $J_x$---separately conserves the two $\mathbb{Z}_2$ symmetries $G_1$ and $G_2$.  The model thus potentially supports a reflection-symmetric cluster state SPT.  One can verify that such a state indeed appears in the phase diagram by examining 
the limit in which the bottom line of Eq.~\eqref{eq:twochain} dominates, with $h \approx J_x$.  Here one can distill the problem to an effective spin-1 model following exactly the same logic presented below Eq.~\eqref{H_triplet}.  In particular, with the aid of Table~\ref{tab.projection} one finds that $H$ maps (modulo a trivial constant) onto $H_{\text{spin-1}}$ from Eq.~\eqref{H_spin1} with $J = J_z$, $J' = J_z'$, and $D = 2(h-J_x)$.  Numerical results from Refs.~\onlinecite{tonegawa1995effect,kaburagi1999spin} then imply that at `large' $h$, the reflection-symmetric cluster state appears for a window of $J_x$ close to $h$, provided $J_z'$ is nonzero.  

\begin{figure}[h]
    \centering
    \includegraphics[width=0.96\linewidth]{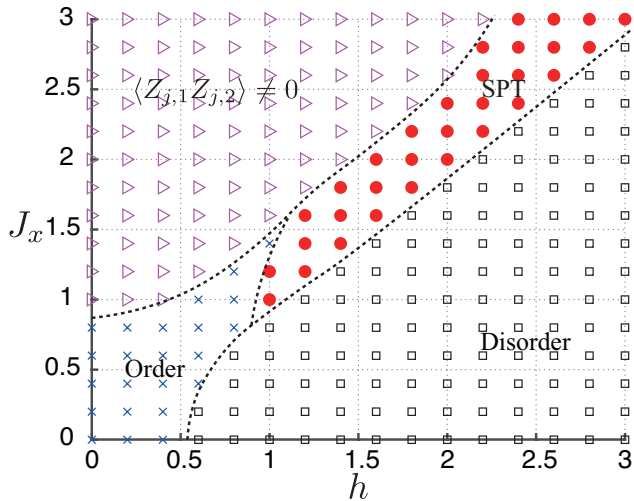}
    \caption{Phase diagram of $H$ in Eq.~\eqref{eq:twochain} obtained using iDMRG (bond dimension 300) with fixed $J_z = 1$ and $J_z' = 0.3$.  The reflection-symmetric cluster-state SPT appears in the region indicated by red dots.  Dashed lines represent rough phase boundaries inferred from the simulated grid points shown.}
    \label{fig:twochainPD}
\end{figure}

To track the reflection-symmetric cluster state in the broader phase diagram---particularly the regime in which the spin-1 mapping breaks down---we simulate the full spin-1/2 ladder model in Eq.~\eqref{eq:twochain} using iDMRG (bond dimension $\chi = 300)$.  Figure~\ref{fig:twochainPD} displays the resulting $(h,J_x)$ phase diagram obtained with $J_z = 1$ and $J_z' = 0.3$.  For relatively small inter-chain coupling $J_x 
\lesssim 0.8$, we find only the familiar disordered and antiferromagnetically ordered phases that persist down to the decoupled-chain limit.  The disordered phase preserves all symmetries and smoothly connects to a trivial product state at $h \rightarrow\infty$.  The ordered state exhibits $\langle Z_{j,y} \rangle \neq 0$ for $y = 1,2$ and thus spontaneously breaks both the $G_1$ and $G_2$ $\mathbb{Z}_2$ symmetries.  Notice that in the $J_x = 0$ decoupled-chain limit, the order-disorder phase transition occurs at $h<1$ due to the frustration-inducing non-zero $J_z'$.

Two new phases emerge at larger $J_x$: First, a partially ordered state with $\langle Z_{j,1} Z_{j,2} \rangle \neq 0$ but $\langle Z_{j,y} \rangle = 0$ eventually supplants the conventional ordered phase.  Intuitively, the `large' $X_{j,1} X_{j,2}$ inter-chain term  anticommutes with $Z_{j,y}$ and thus scrambles antiferromagnetic order in the individual chains, but commutes with $Z_{j,1}Z_{j,2}$ and hence need not suppress the composite order parameter.  Second, we observe the reflection-symmetric cluster-state SPT intervening between the ordered/partially ordered states and the disordered phase.  In our simulations we identify the SPT through string
order parameters akin to Eqs.~\eqref{eq:stringop1} and \eqref{eq:stringop2}:
\begin{align}\label{eq:twochainsptop1}
    \mathcal{O}_1 &= \left\langle Z_{j, 1} \left(\prod_{k=j+1}^{j'-1} X_{k, 2}\right)  Z_{j', 1}\right\rangle 
    \\
    \label{eq:twochainsptop2}
    \mathcal{O}_2 &= \left\langle Z_{j, 2} \left(\prod_{k=j+1}^{j'-1} X_{k, 1}\right)  Z_{j', 2}\right\rangle.
\end{align}
Observe that inter-chain reflection swaps $\mathcal{O}_1 \leftrightarrow \mathcal{O}_2$. In the SPT region of Fig.~\ref{fig:twochainPD} we indeed find $\mathcal{O}_1 = \mathcal{O}_2 \neq 0$ for $|j-j'|\rightarrow \infty$---as required for the cluster state to preserve reflection symmetry.
The SPT resides near the diagonal in Fig.~\ref{fig:twochainPD} where $h \approx J_x$, consistent with expectations from the spin-1 mapping.  We can further solidify the connection to the AKLT chain [and its variant from Eq.~\eqref{H_spin1}] by projecting the above string order parameters into the spin-1 sector using Table~\ref{tab.projection}.  Remarkably, this projection yields, modulo an overall sign, the $x$ and $y$ components of the AKLT-chain SPT order parameters from Eq.~\eqref{AKLTstring},
\begin{equation}
    \mathcal{O}_1 \rightarrow \mathcal{O}^x_{\rm AKLT},~~~\mathcal{O}_2 \rightarrow \mathcal{O}^y_{\rm AKLT}.
\end{equation}

We have also performed DMRG simulations at larger $J_z'$ (see Appendix~\ref{app:numerics}). We find that the minimum $J_x$ value required to stabilize the reflection-symmetric cluster state \emph{decreases} with $J_z'$.
This observation suggests that the reflection-symmetric cluster state should also be analytically accessible as an instability of coupled chains, far from the limit where the spin-1 mapping holds.  With this objective in mind we now revisit the ladder model from the fermionized perspective.

\subsection{Fermionized representation}
\label{fermionized_sec}

\begin{figure}[h]
    \centering
    \includegraphics[width=0.97\linewidth]{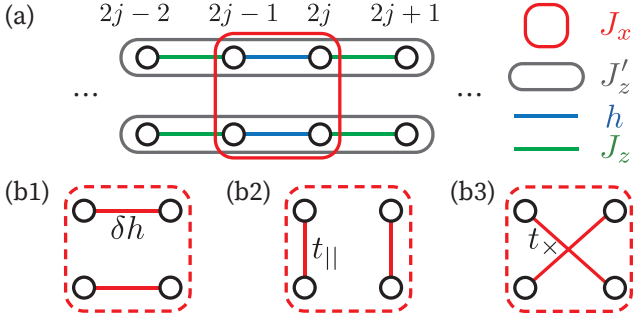}
    \caption{(a) Schematic illustration of the couplings in Eq.~\eqref{H_JW}.  Each circle represents a Majorana fermion. (b) Possible mean-field decouplings of the inter-chain $J_x$ term, leading to (b1) spontaneously broken inter-chain reflection symmetry, (b2) partial order with $\langle Z_{j,1}Z_{j,2}\rangle \neq 0$, and (b3) the reflection-symmetric cluster state SPT.}
    \label{fig:decoupling}
\end{figure}

We fermionize Eq.~\eqref{eq:twochain} via the Jordan-Wigner transformation
\begin{equation}
\begin{aligned}
    &\gamma_{2j-1, 1} = Z_{j, 1}\prod_{k=1}^{j-1}X_{k, 1}, \quad \gamma_{2j, 1} = iX_{j, 1} \gamma_{2j-1, 1},\\
    &\gamma_{2j-1, 2} = G_1 Z_{j, 2}\prod_{k=1}^{j-1}X_{k, 2} , \quad \gamma_{2j, 2} = iX_{j,2} \gamma_{2j-1, 2};
\end{aligned}
\end{equation}
on the second line, the factor of $G_1$ ensures that Majorana fermions $\gamma_{j,1}$ and $\gamma_{j',2}$ on different chains anticommute.  We then arrive at an equivalent fermion model that is local (due to the separately conserved $\mathbb{Z}_2$ symmetries for each chain),
\begin{align}
    H&= \sum_{j}\bigg{\{}\sum_{y=1,2}  \big{[}i J_z \gamma_{2j, y}\gamma_{2j+1, y}-ih\gamma_{2j-1, y}\gamma_{2j, y}   
    \nonumber \\
    &~~~~~~~~~~+J_z' (i\gamma_{2j, y} \gamma_{2j+1,y})(i\gamma_{2j+2,y} \gamma_{2j+3,y})\big{]}
    \nonumber \\
    &~~~~~~~~~~+J_x (i\gamma_{2j-1, 1} \gamma_{2j,1})(i \gamma_{2j-1,2} \gamma_{2j,2})\bigg{\}}.
    \label{H_JW}
\end{align}
Above we have been cavalier about limits on the sums to simplify the presentation.  \sfigure{fig:decoupling}{a} illustrates the set of fermion couplings encoded in Eq.~\eqref{H_JW}.

Next, we will explore phases that inter-chain coupling promotes by considering three possible mean-field decouplings of the fermionized $J_x$ term above; see Figs.~\hyperref[fig:decoupling]{\ref{fig:decoupling}(b1-b3)}.  For this exercise it will prove useful to express correlators indicating partial $\langle Z_{j,1} Z_{j,2}\rangle$ order and SPT order in terms of Majorana fermions as follows:
\begin{align}
    &\left|\langle (Z_{j,1} Z_{j,2})(Z_{j',1} Z_{j',2})\rangle\right| = \left| \left\langle 
    \begin{smallmatrix}
     \gamma_{2j} & \gamma_{2j+1} & \cdots & \gamma_{2j'-2} & \gamma_{2j'-1}\\
    \gamma_{2j} & \gamma_{2j+1} & \cdots & \gamma_{2j'-2} & \gamma_{2j'-1}\\
    \end{smallmatrix}
    \right\rangle \right|,
    \label{ZZZZ}
    \\
    &~~~~~~|\mathcal{O}_1| = 
    \left| \left\langle 
    \begin{smallmatrix}
     \gamma_{2j} & \gamma_{2j+1} & \cdots & \gamma_{2j'-2} & \gamma_{2j'-1}\\
    & \gamma_{2j+1} & \cdots & \gamma_{2j'-2} & \\
    \end{smallmatrix}
    \right\rangle \right|,
    \label{O1ferm}
    \\
    &~~~~~~|\mathcal{O}_2| = \left| \left\langle 
    \begin{smallmatrix}
     & \gamma_{2j+1} & \cdots & \gamma_{2j'-2} &\\
    \gamma_{2j} & \gamma_{2j+1} & \cdots & \gamma_{2j'-2} & \gamma_{2j'-1}\\
    \end{smallmatrix}
    \right\rangle \right|.
    \label{O2ferm}
\end{align}
On the right sides, we indicate the Majorana operators that are multiplied to give the order parameters on the left; the top and bottom rows respectively correspond to operators living on the upper and lower chains of the ladder.   Absolute values are included merely to avoid tracking factors of $i$.  

The most trivial decomposition of the $J_x$ term, sketched in \sfigref{fig:decoupling}{b1}, follows from the replacement
\begin{align}
    &J_x(i\gamma_{2j-1, 1} \gamma_{2j,1})(i \gamma_{2j-1,2} \gamma_{2j,2}) \rightarrow 
    \nonumber \\
    &~~~~~~\delta h(i \gamma_{2j-1,2} \gamma_{2j,2} - i\gamma_{2j-1, 1} \gamma_{2j,1}).
\end{align}
Here $\delta h = J_x \langle i\gamma_{2j-1, 1} \gamma_{2j,1} \rangle = - J_x\langle i\gamma_{2j-1, 2} \gamma_{2j,2} \rangle$ represents an opposite-sign shift in the transverse field for the two chains (recall that we assume $J_x \geq 0$).  A state characterized by $\delta h \neq 0$ would spontaneously break inter-chain reflection symmetry in a way that promotes order in one chain and disorder in the other---which we do not observe in the phase diagram from Fig.~\ref{fig:twochainPD}.  

The remaining two $J_x$ decouplings generate different patterns of spontaneous inter-chain fermion tunneling.  \sfigure{fig:decoupling}{b2} corresponds to the decoupling
\begin{align}
    &J_x(i\gamma_{2j-1, 1} \gamma_{2j,1})(i \gamma_{2j-1,2} \gamma_{2j,2}) \rightarrow 
    \nonumber \\
    &~~~~~~ -t_{||}(i \gamma_{2j-1,1} \gamma_{2j-1,2} + i\gamma_{2j, 1} \gamma_{2j,2})
    \label{t||}
\end{align}
with `vertical' hopping amplitude $t_{||} = J_x \langle i \gamma_{j,1} \gamma_{j,2}\rangle$.  From the right side of Eq.~\eqref{ZZZZ}, we see that the special limit with $t_{||} = J_x$ yields the partially ordered $\langle Z_{j,1} Z_{j,2}\rangle \neq 0$ phase captured by DMRG. 
[By contrast, the `dangling' $\gamma_{2j}, \gamma_{2j'-1}$ Majorana operators in Eqs.~\eqref{O1ferm} and \eqref{O2ferm} imply that the string order parameters $\mathcal{O}_{1,2}$ vanish.] 
Twofold ground-state degeneracy of the partially ordered phase is encoded in this representation through the two (degenerate) choices for the sign of $t_{||}$.  

Finally, \sfigref{fig:decoupling}{b3} represents the decoupling
\begin{align}
    &J_x(i\gamma_{2j-1, 1} \gamma_{2j,1})(i \gamma_{2j-1,2} \gamma_{2j,2}) \rightarrow 
    \nonumber \\
    &~~~~~~ -t_{\times}(i \gamma_{2j-1,1} \gamma_{2j,2} + i\gamma_{2j-1, 2} \gamma_{2j,1}),
\end{align}
where $t_{\times} = J_x \langle i \gamma_{2j-1,1} \gamma_{2j,2}\rangle = J_x \langle i \gamma_{2j-1,2}\gamma_{2j,1}\rangle$ encodes a `crossed' inter-chain fermion tunneling amplitude.  The pattern of inter-chain fermion hopping generated at $t_{\times} \neq 0$ flips the situation compared to Eq.~\eqref{t||}: In the extreme limit where $t_\times = J_x$, at $|j-j'| \rightarrow \infty$  Eq.~\eqref{ZZZZ} now vanishes while Eqs.~\eqref{O1ferm} and \eqref{O2ferm} are non-zero---indicating reflection-symmetric cluster-state SPT order.  Ignoring the $J_z'$ term for simplicity then yields a minimal mean-field Hamiltonian 
\begin{align}
    H_{\rm MF}&= \sum_{j}\bigg{[}\sum_{y=1,2}  \big{(}i J_z \gamma_{2j, y}\gamma_{2j+1, y}-ih\gamma_{2j-1, y}\gamma_{2j, y}\big)   
    \nonumber \\
    &~~~~~~~~~~-t_{\times}(i \gamma_{2j-1,1} \gamma_{2j,2} + i\gamma_{2j-1, 2} \gamma_{2j,1})\bigg{]}
    \label{H_MF}
\end{align}
for the SPT.

We stress that $J_z'$ is crucial for energetically favoring the decoupling used here, but is not essential for understanding universal properties of the resulting phase.  Note also that, when writing Eq.~\eqref{H_MF}, we neglected possible renormalization of $J_z$ and $h$ by $J_x$.  One can show (e.g., by studying the spectrum of $H_{\rm MF}$) that when $\left| |J_z| - |h| \right| < |t_{\times}| < |J_z| + |h|$, the mean-field Hamiltonian hosts a single unpaired Majorana zero mode on each end, similar to the topological phase of a Kitaev chain~\cite{Kitaev2001}.  The two edge Majorana modes together with the arbitrary sign of $t_\times$ encode the fourfold degeneracy characteristic of the reflection-symmetric cluster state.  As an aside, the preceding analysis provides an explicit microscopic counterpart of the field-theoretic coupled-critical-chain construction explored in the context of Rydberg arrays in Ref.~\onlinecite{Slagle22}.

\begin{figure}[h]
    \centering
    \includegraphics[width=0.97\linewidth]{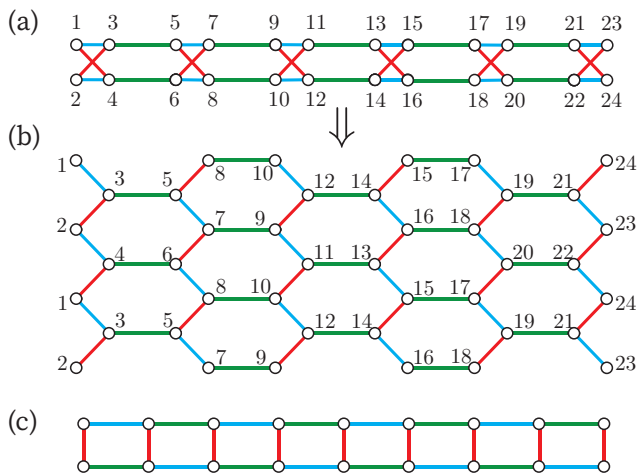}
    \caption{(a) Schematic representation of the two-chain Hamiltonian from Eq.~\eqref{H_MF} that provides a mean-field description of the reflection-symmetric cluster state SPT.  (b) Equivalent unfolded representation of (a) that reveals a connection to the Kitaev honeycomb model.  In particular, a Kitaev honeycomb strip on the same geometry also realizes an SPT when the ground state resides in the flux sector corresponding to Eq.~\eqref{H_MF}.  (c) Alternative Kitaev honeycomb strip geometry studied in Refs.~\onlinecite{Catuneanu19,DeGottardi_2011}, which for comparison does not realize an SPT.  
    }
    \label{fig:unfold}
\end{figure}

Interestingly, the form of $H_{\rm MF}$ already hints at a connection to the Kitaev honeycomb model.  Figure~\ref{fig:unfold}(a) sketches the couplings in $H_{\rm MF}$ for a system with a total of 24 Majorana fermions, labeled in a way that is convenient for the present aim.  Figure~\ref{fig:unfold}(b) presents an equivalent `unfolded' version---which represents nearest-neighbor-hybridized Majorana fermions on a honeycomb strip wrapped into a cylinder.  Correspondingly, $H_{\rm MF}$ has an identical spectrum to the Kitaev honeycomb model defined on the same cylindrical strip, in the sector with $\pi$ flux per plaquette \footnote{Here we refer specifically to the flux in the $\mathbb{Z}_2$ gauge field that the itinerant Majorana fermions couple to, i.e., $\pi$ flux corresponds to plaquette operator $W_p := XYZXYZ = +1$.}.
It follows that a spin-1/2 system on the strip in \sfigref{fig:unfold}{a} with $XX$, $YY$ and $ZZ$ couplings on blue, green and red bonds is also an SPT in the presence of perturbations that stabilize the $\pi$-flux sector (see Appendix~\ref{app:kitaevstrip} for further discussion).  We should contrast this honeycomb strip geometry to the more well-studied Kitaev strip in \sfigref{fig:unfold}{c}~\cite{Catuneanu19,DeGottardi_2011}---which instead realizes symmetry-breaking order in the analogous regime as also discussed in Appendix~\ref{app:kitaevstrip}. 

\subsection{Finite-size hybridization of edge zero modes}

The reflection-symmetric cluster state SPT emerging from Eq.~\eqref{eq:twochain} exhibits a finite correlation length $\xi$.  Consequently, on any system of finite size $L$, the anomalous edge spin-1/2's generically hybridize and move away from zero energy (by an amount that becomes exponentially small in $L/\xi$ when $L \gg \xi$).  The symmetries listed in Table~\ref{tab.edgetransform}  restrict the residual interaction between edge modes localized on the left and right ends of the ladder to the form
\begin{equation}\label{eq:Hedge}
     H_{\rm edge} = \Gamma_z \left(\mathcal{X}_L \mathcal{Y}_R + \mathcal{Y}_L \mathcal{X}_R\right) + \mathcal{J}_z \mathcal{Z}_L \mathcal{Z}_R.
\end{equation}
The couplings $\Gamma_z$ and $\mathcal{J}_z$ are non-universal and depend on details of the microscopic Hamiltonian generating the SPT.  For a given set of microscopic parameters, one can numerically extract $\Gamma_z$ and $\mathcal{J}_z$ by matching the eigenvalues of $H_{\rm edge}$ (given by $\{-\mathcal{J}_z, -\mathcal{J}_z, -2\Gamma_z+\mathcal{J}_z, 2\Gamma_z+\mathcal{J}_z\}$) with the four lowest energies extracted from simulations of the microscopic two-chain model with open boundary conditions.

\begin{figure}[h]
    \centering
    \includegraphics[width=\linewidth]{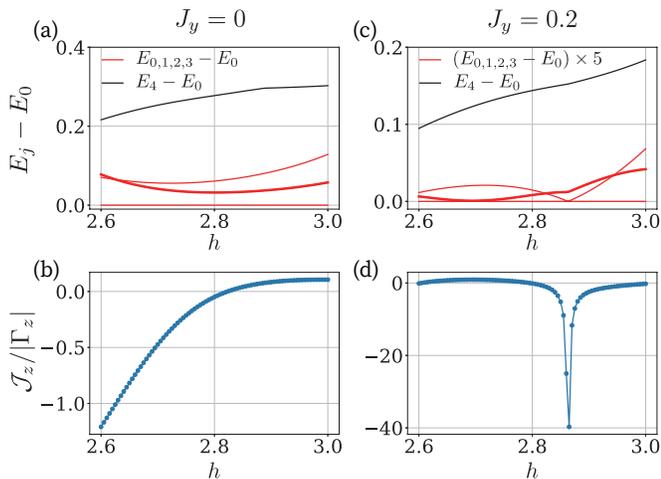}
    \caption{(a,c) Low-lying energy levels and (b,d) corresponding $\mathcal{J}_z/|\Gamma_z|$ values for an $N = 8$ chain described by Eq.~\eqref{eq:twochain} (left panels), and with an additional intra-chain $YY$ interaction from Eq.~\eqref{deltaH} (right panels).  In all panels we fix $(J_z, J_z', J_x)=(1, 0.3, 3)$ and vary the transverse field $h$ within the SPT regime; for the right panels we take $J_y = 0.2$.  Red levels in (a,c) correspond to the four lowest energies $E_{0,1,2,3}$---emerging from SPT edge modes that hybridize due to finite-size effects---measured relative to the ground state energy $E_0$; the thick red lines are doubly degenerate.  Note that in (c) the red curves are scaled by a factor of 5 for visibility.  Black curves represent the next excited energy, i.e., the bulk gap of the SPT phase. The pronounced dip in (d) corresponds to a regime for which $\mathcal{J}_z$ dominates over $\Gamma_z$. }
    \label{fig:YYterm}
\end{figure}

The left panels of Fig.~\ref{fig:YYterm} illustrate (a) the low-lying level structure and (b) the extracted $\mathcal{J}_z/|\Gamma_z|$ values for an $N = 8$ chain described by Eq.~\eqref{eq:twochain} with $J_z = 1, J_z' = 0.3$, and $J_x = 3$; the horizontal axes cover an interval of the transverse field $h$ corresponding to the SPT phase in Fig.~\ref{fig:twochainPD}.  For these microscopic parameters we obtain $|\mathcal{J}_z| \lesssim |\Gamma_z|$, i.e., the $\Gamma_z$ term tends to dominate.  To underscore the non-universality of this result, Fig.~\ref{fig:YYterm}(c,d) presents the same quantities as (a,b) but incorporating a new intra-chain term 
\begin{equation}
     \delta H = J_y\sum_{y = 1,2} \sum_{j = 1}^{N-1} Y_{j,y}Y_{j+1,y}
     \label{deltaH}
\end{equation}
with $J_y = 0.2$ (all other parameters are unchanged).  The $YY$ term preserves the $\mathbb{Z}_2 \times \mathbb{Z}_2$ symmetry as well as inter-chain reflection symmetry and thus need not disrupt the SPT phase; in fact the bulk gap increases significantly in (c) relative to (a), indicating that $J_y = 0.2$ actually strengthens the SPT.  Moreover, panel (d) reveals a parameter regime in which $\mathcal{J}_z/|\Gamma_z|$ \emph{diverges}---establishing proof-of-concept that $\mathcal{J}_z$ can dominate over $\Gamma_z$.  For rough intuition, one can view the $YY$ term above as enabling non-trivial $2k_F$ oscillations that modulate the overlap of Majorana end states described at the mean-field level by Eq.~\eqref{H_MF}.
A finite-size SPT with $|\mathcal{J}_z| \gg |\Gamma_z|$ will be of particular interest in the next section because it can be used as a building block of a Kitaev honeycomb quantum spin liquid.

\section{Kitaev honeycomb model variant from SPT arrays}
\label{KitaevVariant}

Once we know how to construct a reflection-symmetric cluster state, we can use finite-sized SPT blocks to assemble a variant of the Kitaev honeycomb model that exhibits only subextensive conserved quantities.  Consider the setup in Fig.~\ref{fig:array} that features $\mathcal{N}$ spin-1/2 chains (horizontal dashed lines), each with a spin-flip $\mathbb{Z}_2$ symmetry.  We incorporate intra- and inter-chain couplings that form SPT blocks in the pattern shown in the figure---yielding a new set of low-energy degrees of freedom consisting of anomalous spin-1/2 boundary modes (grey circles with arrows) arranged on an anisotropic honeycomb lattice.  Below we describe the edge spin at position $j$ with operators $\mathcal{X}_j$, $\mathcal{Y}_j$, and $\mathcal{Z}_j$ and discard the now redundant $L/R$ subscript used previously to label the two ends of an SPT.

Recalling the transformation properties from Table~\ref{tab.edgetransform}, the $\mathbb{Z}_2$ symmetries associated with each chain heavily constrain the allowed two-spin interactions between anomalous edge spins in the array.  A given pair of edge spins can indeed only hybridize if the SPT's to which they belong share at least one chain.  Further assuming that each edge spin only interacts with its partner on the same SPT and its two nearest-neighbors from adjacent SPT's, we obtain the pattern of couplings illustrated by green, blue, and red bonds in Fig.~\ref{fig:array}.  Specifically, green bonds only allow $\mathcal{X}\mathcal{X}$-type interactions; blue bonds only allow $\mathcal{Y}\mathcal{Y}$-type interactions; and red bonds allow couplings of the form in Eq.~\eqref{eq:Hedge}.  The effective Hamiltonian for the array thus becomes
\begin{align}\label{eq:array}
    H_{\rm array} &= \mathcal{J}\left(\sum_{\rm{green}} \mathcal{X}_j \mathcal{X}_{j'} + \sum_{\rm{blue}} \mathcal{Y}_j \mathcal{Y}_{j'}\right)
    \\ \nonumber
    &+ \sum_{\rm{red}} \left[\mathcal{J}_z \mathcal{Z}_j \mathcal{Z}_{j'} + \Gamma_z (\mathcal{X}_j \mathcal{Y}_{j'} + \mathcal{Y}_{j} \mathcal{X}_{j'})\right],
\end{align}
where the colors indicate the $j,j'$ bonds summed over and we assumed reflection symmetry that equates the $\mathcal{X}\mathcal{X}$ and $\mathcal{Y}\mathcal{Y}$ couplings. Remarkably, downgrading the number of conserved quantities from extensive to subextensive generates only a single two-spin nearest-neighbor perturbation ($\Gamma_z$) to the exactly solvable Kitaev honeycomb model!

\begin{figure}[h]
    \centering
    \includegraphics[width=\linewidth]{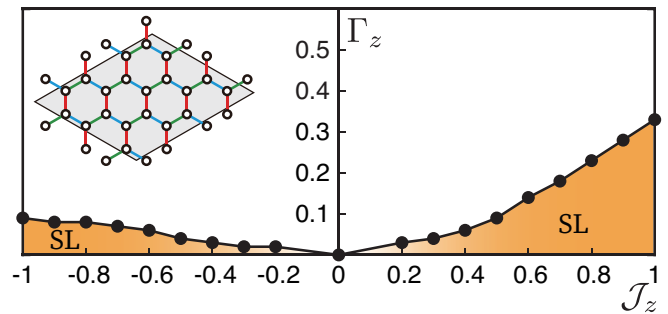}
    \caption{Phase diagram of the SPT array model in Eq.~\eqref{eq:array} obtained from exact diagonalization of the 20-site cluster shown in the inset.  Simulations assume periodic boundary conditions with $\mathcal{J} = 1$.  The phase boundary of the spin liquid (SL) is determined from maxima of $-\partial^2 E_0 / \partial \Gamma_z^2$, where $E_0$ is the ground-state energy.   }
    \label{fig:c1c2phasediagram}
\end{figure}

Previous numerical studies have explored the phase diagram of the Kitaev honeycomb model perturbed by various couplings including Heisenberg terms and off-diagonal exchange anisotropies~\cite{Chaloupka2010,Jiang2011,Chaloupka2013,Rau2014,Gotfryd2017,Gordon2019,Lee2020,Yang2020,Zhang2021,Luo2021,Nanda2021,yang2022counterrotating}.  Our array Hamiltonian, however, includes a spatially anisotropic $\Gamma_z$ term that has not been simulated to our knowledge.  We therefore obtain the phase diagram of Eq.~\eqref{eq:array} using exact diagonalization of the 20-site cluster shown in the inset of Fig.~\ref{fig:c1c2phasediagram}, assuming periodic boundary conditions and fixing $\mathcal{J} = 1$.  As a baseline, for $\Gamma_z \neq 0$ but $\mathcal{J}_z = 0$ the array realizes a trivial phase in which the SPT blocks are completely disentangled from one another.  Conversely, for $\Gamma_z = 0$ but $\mathcal{J}_z \neq 0$ the gapless spin liquid phase of the Kitaev honeycomb model emerges. We determine the stability of the spin liquid at finite $\Gamma_z$ by examining extrema of the second derivative of the ground state energy---yielding the phase diagram in Fig.~\ref{fig:c1c2phasediagram}. For non-zero $\mathcal{J}_z$, the spin liquid persists over a finite window in $\Gamma_z$, with antiferromagnetic $\mathcal{J}_z$ offering more resilience than ferromagnetic $\mathcal{J}_z$ (similar to trends observed in other studies \cite{Rau2014,Yang2020,Zhang2021,Nanda2021}). The key takeaway is that in the regime $|\mathcal{J}_z| \gg |\Gamma_z|$---which can be satisfied, e.g., by adding $YY$ terms to the model in Eq.~\eqref{eq:twochain}---the SPT array indeed realizes a gapless Kitaev spin liquid.

\section{Discussion}
\label{Discussion}

We introduced microscopic realizations of a cluster-state-like SPT for a spin-1/2 ladder enjoying inter-chain reflection symmetry along with a pair of subsystem $\mathbb{Z}_2$ symmetries associated with globally flipping all spins on a given chain.  Reflection symmetry in particular distinguishes this phase from the canonical cluster state described by the commuting-projector Hamiltonian in Eq.~\eqref{Hcluster}.  As a conceptually simple microscopic route, we first considered the limit wherein the spin-1/2 ladder reduces to a single spin-1 chain, due to a fine-tuned interplay between transverse-field and inter-chain coupling terms [Eq.~\eqref{H_triplet}] that energetically favor three out of the four available states for a given rung.  We leveraged the frustration-free AKLT Hamiltonian to establish that the reflection-symmetric cluster state SPT for the spin-1/2 ladder corresponds to the Haldane phase for the effective spin-1 chain.  Guided by numerical results on an alternative spin-1 model [Eq.~\eqref{H_spin1}], we further established via DMRG that the reflection-symmetric cluster state persists in relatively simple spin-1/2 ladder models---well away from the limit where the spin-1 mapping applies---that invoke only one- and two-spin interactions; see Eqs.~\eqref{eq:twochain} and \eqref{deltaH}.  

The hallmark anomalous edge spin-1/2's for the reflection-symmetric cluster state furnish very natural degrees of freedom for constructing a minimally perturbed Kitaev honeycomb model.  One can usefully associate the $\mathcal{X}$ and $\mathcal{Y}$ components of the anomalous edge spins with opposite chains in the ladder, in the following sense: $\mathcal{X}$ changes sign under the spin-flip symmetry for one chain, while $\mathcal{Y}$ changes sign under the spin-flip symmetry for the other chain (recall Table~\ref{tab.edgetransform}).  It follows that in the SPT array from Fig.~\ref{fig:array}, the \emph{only} symmetry-allowed interactions among anomalous edge spins connected by the green and blue bonds correspond precisely to the $\mathcal{X}\mathcal{X}$ and $\mathcal{Y}\mathcal{Y}$ interactions built into the Kitaev honeycomb model.  Moreover, symmetry under reflections about any of the red bonds in Fig.~\ref{fig:array} constrains the  $\mathcal{X}\mathcal{X}$ and $\mathcal{Y}\mathcal{Y}$ couplings to be equal.  Interactions among the anomalous edge spins within a given SPT block are, however, less constrained: Along the red bonds, symmetry allows the $\mathcal{Z}\mathcal{Z}$ interactions that complete the Kitaev honeycomb model, along with an additional off-diagonal exchange anisotropy [$\Gamma_z$ term in Eq.~\eqref{eq:array}].  The $\Gamma_z$ term mediates restrictive dynamics for $\mathbb{Z}_2$ fluxes that are completely static in the pure Kitaev limit; namely, fluxes can tunnel in the vertical direction of Fig.~\ref{fig:array}, but not the `diagonal' directions, which is the price paid for invoking subextensive conserved quantities.   We showed explicitly that the gapless spin liquid phase survives a finite threshold of $\Gamma_z$, and demonstrated that a concrete spin-1/2 ladder model can indeed give rise to $\Gamma_z$ values well below that threshold.  

These results highlight the potential power of exploring Hamiltonians designed to emulate exotic ground-states of exactly solvable models, but with only a subextensive number of conserved quantities.  Indeed, as proof of concept, our construction rigorously establishes that Kitaev honeycomb model phenomenology can arise from an entirely different microscopic framework built upon interactions needed to stabilize a two-chain SPT.  We do not expect that the array from Fig.~\ref{fig:array}, given the large unit cell, will find direct experimental relevance in solid-state systems.  Nevertheless, we hope that our study will motivate the development of related models that do expand the landscape of candidate spin liquid materials.  

Perhaps the most immediate experimentally relevant implication of our study concerns the reflection-symmetric cluster state for a single two-chain system.  Throughout we enforced \emph{three} $\mathbb{Z}_2$ symmetries---inter-chain reflection, plus a pair of spin-flip symmetries.  These symmetries are essential for maintaining the structure of our perturbed Kitaev honeycomb model in the SPT array; however,  only two $\mathbb{Z}_2$'s are strictly required to maintain a nontrivial SPT for an elementary two-chain block.  In particular, it suffices to retain inter-chain reflection while relaxing the subsystem spin-flip symmetries into a more generic \emph{global} spin-flip symmetry.  Inspection of Table~\ref{tab.edgetransform} reveals that the edge spin-1/2's indeed remain protected zero-energy degrees of freedom when enforcing only $G_{\rm ref}$ and $G_1\times G_2$.  Finding experimental realizations for the reflection-symmetric cluster state with these realistic symmetries poses an interesting challenge for future research.

%%%%%%%%%%%%%%%%%%%%%%%%%%%%%%%%%%%%%%%%%%%%%%%%
\begin{acknowledgments}

It is a pleasure to acknowledge illuminating conversations with Xie Chen and Gabor Halasz.  The U.S. Department of
Energy, Office of Science, National Quantum Information Science Research Centers, Quantum Science Center supported the construction and numerical analysis of the models studied in this paper. Additional support was provided by the Caltech Institute for Quantum
Information and Matter, an NSF Physics Frontiers Center with support of the Gordon and Betty Moore Foundation through Grant GBMF1250, and the Walter Burke
Institute for Theoretical Physics at Caltech. DFM was supported by the Israel Science Foundation (ISF) under grant 2572/21. 

\end{acknowledgments}

\appendix

\section{AKLT model and its edge modes}
\label{AKLTreview}

\begin{figure}[h]
    \centering
    \includegraphics[width=\linewidth]{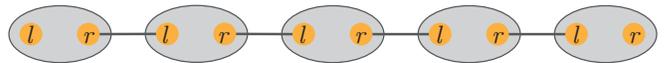}
    \caption{Illustration of the AKLT model ground state. Spin-1 degrees of freedom, depicted by ovals, are decomposed into left ($l$) and right ($r$) spin-1/2's. The spin-1/2's connected by solid lines form a singlet state $\frac{\ket{\uparrow\downarrow}-\ket{\downarrow\uparrow}}{\sqrt{2}}$, while the two spin-1/2's in each oval are projected into the triplet subspace formed by $\ket{\uparrow\uparrow}$, $\ket{\downarrow\downarrow}$, $\frac{\ket{\uparrow\downarrow}+\ket{\downarrow\uparrow}}{\sqrt{2}}$. Dangling spin-1/2's at the left and right edges form anomalous edge states for the spin-1 chain.}
    \label{fig:aklt}
\end{figure}

In this appendix we review the spin-1 AKLT model and discuss the transformation of its edge operators under different symmetries. The AKLT Hamiltonian,
\begin{equation}
    H_{\rm AKLT} = \sum_j\left[\vec{S}_j \cdot \vec{S}_{j+1}+\frac{1}{3}\left(\vec{S}_j \cdot \vec{S}_{j+1}\right)^2\right],
\end{equation}
represents a sum of projection operators of neighboring spin-1's to the total spin-2 sector. Therefore, for a given nearest neighbor pair, states in the total spin-2 sector uniquely incur an energy penalty, so that the energy is minimized when every such pair resides in the sector with total spin 0 or 1.  One can efficiently satisfy this condition by first decomposing the spin-1 on site $j$ into two spin-1/2's denoted by $\ket{s^l_j s^r_j}$ with $s=\uparrow,\downarrow$, then putting the `$r$' spin 1/2 from site $j$ and `$l$' spin 1/2 from site $j+1$ into a singlet, and finally projecting all sites into the physical spin-1 subspace; see Fig.~\ref{fig:aklt} for an illustration.  Neglecting edge effects for now, the ground state so constructed takes the valence bond form
\begin{equation}
    \ket{\psi_{\rm AKLT}} = \prod_j P_j \prod_j \frac{\ket{\uparrow^r_j \downarrow^l_{j+1}} - \ket{\downarrow^r_j \uparrow^l_{j+1}}}{\sqrt{2}}
\end{equation}
where
\begin{equation}
    P_j = \ket{1_j}\bra{\uparrow^l_j \uparrow^r_j} + \ket{0_j}\frac{\bra{\uparrow^l_j \downarrow^r_j} + \bra{\downarrow^l_j \uparrow^r_j}}{\sqrt{2}} + \ket{-1_j}\bra{\downarrow^l_j \downarrow^r_j}
\end{equation}
imposes spin-1 projection.  

For an open chain in the thermodynamic limit, the leftmost and rightmost `dangling' spin-1/2's in Fig.~\ref{fig:aklt} decouple and can orient in any direction without changing the system's energy, giving rise to fourfold degeneracy. The ground states can be labeled by the edge spin-1/2 configurations:
\begin{equation}
    \ket{s_L \cdots s_R} = \prod_{j=1}^N P_j \ket{s_L}\prod_{j=1}^{N-1} \frac{\ket{\uparrow^r_j \downarrow^l_{j+1}} - \ket{\downarrow^r_j \uparrow^l_{j+1}}}{\sqrt{2}} \ket{s_R},
\end{equation}
where $s_{L,R}=\uparrow,\downarrow$ denote the state of the left and right edge modes.
We define edge operators $O_{L,R}$ that act on the edge spin $s_{L,R}$ as
\begin{align}
    O_L \ket{s_L \cdots s_R} &\equiv \ket{(O_L s_L) \cdots s_R}, \nonumber \\
    O_R \ket{s_L \cdots s_R} &\equiv \ket{s_L \cdots (O_R s_R)},
\end{align}
with $O=X,Y,Z$ denoting various Pauli operators. As an example, $X_L\ket{\uparrow \cdots \uparrow} = \ket{\downarrow \cdots \uparrow}$, $Y_R\ket{\uparrow \cdots \uparrow} = i \ket{\uparrow \cdots \downarrow}$, etc.

To understand the transformation of edge operators under symmetries---specifically Eqs.~\eqref{G12} and \eqref{Gref}---we first write down an important property of the projection operator,
\begin{equation}
    -e^{i \pi \vec{n}\cdot \vec{S}_j} P_j = P_j (\Vec{n}\cdot \vec{\sigma}^l_j)\otimes(\Vec{n}\cdot\vec{\sigma}^r_j).
    \label{relation}
\end{equation}
Here $\Vec{S}_j = (S^x_j, S^y_j, S^z_j)$ is the spin-1 operator for site $j$, $\Vec{\sigma}^{l,r}_j=(X^{l,r}_j, Y^{l,r}_j, Z^{l,r}_j)$ is the Pauli operator acting on the left (right) spin-1/2 on site $j$, and $\Vec{n}=(n_x, n_y, n_z)$ is any unit vector satisfying $n_x^2+n_y^2+n_z^2=1$. Equation~\eqref{relation} links the rotation of a spin-1 to simultaneous rotation of two spin-1/2's in the spin-1/2 representation.

Thus, we have
\begin{equation}
\begin{aligned}
    &\prod_{j=1}^N \left(-e^{i \pi \vec{n}\cdot \vec{S}_j}\right) \ket{s_L\cdots s_R}\\
    =&\prod_{j=1}^N \left(-e^{i \pi \vec{n}\cdot \vec{S}_j} P_j\right) \ket{s_L}\prod_{j=1}^{N-1} \frac{\ket{\uparrow^r_j \downarrow^l_{j+1}} - \ket{\downarrow^r_j \uparrow^l_{j+1}}}{\sqrt{2}} \ket{s_R}\\
    =&\prod_{j=1}^N \left[P_j (\Vec{n}\cdot \vec{\sigma}^l_j)\otimes(\Vec{n}\cdot\vec{\sigma}^r_j)\right] \ket{s_L}\prod_{j=1}^{N-1} \frac{\ket{\uparrow^r_j \downarrow^l_{j+1}} - \ket{\downarrow^r_j \uparrow^l_{j+1}}}{\sqrt{2}} \ket{s_R}\\
    =&\prod_{j=1}^N P_j \left[ (\Vec{n}\cdot \vec{\sigma}_L) \ket{s_L} \right] \prod_{j=1}^{N-1} \frac{\ket{\uparrow^r_j \downarrow^l_{j+1}} - \ket{\downarrow^r_j \uparrow^l_{j+1}}}{\sqrt{2}} \left[ (\Vec{n}\cdot \vec{\sigma}_R) \ket{s_R} \right]\\
    =&(-1)^{N-1}\ket{(\Vec{n}\cdot \vec{\sigma}_L) s_L  \cdots (\Vec{n}\cdot \vec{\sigma}_R) s_R }.
\end{aligned}
\end{equation}
From the third to fourth line, we used the fact that $(\Vec{n}\cdot \vec{\sigma}^r_j)\otimes(\Vec{n}\cdot\vec{\sigma}^l_{j+1})$ gives a factor of $(-1)$ when acting on the singlet state $\frac{\ket{\uparrow^r_j \downarrow^l_{j+1}} - \ket{\downarrow^r_j \uparrow^l_{j+1}}}{\sqrt{2}}$, and $\Vec{\sigma}_{L,R}$ denotes the Pauli operator on the left or right edge spin-1/2.
Therefore the overall rotation of the spin-1 chain, $\prod_{j=1}^N e^{i \pi \vec{n}\cdot \vec{S}_j}$, is equivalent to applying corresponding Pauli operator $\Vec{n}\cdot\Vec{\sigma}$ to the edge modes in the spin-1/2 representation. As an example, we have $G_2 \ket{s_L \cdots s_R} = (-1)^{N-1}\ket{(X_L s_L) \cdots (X_R s_R)}$, etc. We can then read off the transformation of edge operators under symmetries,
\begin{equation}
    \begin{aligned}
        G_1 X_{L,R} = -X_{L,R} G_1, &\quad G_1 Y_{L,R} = Y_{L,R} G_1, \\
        G_2 X_{L,R} = X_{L,R} G_2, &\quad G_2 Y_{L,R} = -Y_{L,R} G_2, \\
        G_{\rm ref} X_{L,R} = Y_{L,R} G_{\rm ref}, &\quad G_{\rm ref} Y_{L,R} = X_{L,R} G_{\rm ref}.\\
    \end{aligned}
\end{equation}
To arrive at the edge operators used in the main text, one swaps the definition of $X$ and $Y$ operators on the right edge, $X_R \leftrightarrow Y_R$, so that $Y_R$ is odd under $G_1$ while $X_R$ is odd under $G_2$, and then invokes calligraphic fonts via $X \rightarrow \mathcal{X}, Y \rightarrow \mathcal{Y}, Z \rightarrow \mathcal{Z}$.  In this way one arrives at the transformation of edge operators given in Table~\ref{tab.edgetransform}.

\section{Generating the AKLT model from a spin-1/2 ladder}
\label{app.akltfrom2chain}
The AKLT model can be realized by projecting a spin-1/2 ladder to a spin-1 chain using Eq.~\eqref{H_triplet} and then adding appropriate interactions. Specifically, we find that under projection,
\begin{equation}
    \begin{aligned}
        &h_j\equiv\sum_{y=1,2} \left[\frac{1}{6} X_{j,y} X_{j+1,y} + \frac{1}{6} Y_{j,y} Y_{j+1,y} + \frac{5}{6} Z_{j,y} Z_{j+1,y}\right]\\
        &~~~ + \frac{1}{12} \left[(X_{j,1}-1) (X_{j+1,2}-1) + (X_{j,2}-1) (X_{j+1,1}-1)\right] \\
        &~~~ +\frac{1}{6} Z_{j,1} Z_{j,2} Z_{j+1,1} Z_{j+1,2}\\
        &~~~ +\frac{5}{12} \left(  Y_{j,1} Z_{j,2} Y_{j+1,1} Z_{j+1,2} +  Z_{j,1} Y_{j,2} Z_{j+1,1} Y_{j+1,2} \right) 
    \end{aligned}
\end{equation}
maps (up to a constant) to the AKLT Hamiltonian term $\vec{S}_j \cdot \vec{S}_{j+1}+\frac{1}{3}\left(\vec{S}_j \cdot \vec{S}_{j+1}\right)^2$. Thus, $H_{\text{triplet}}+\sum_j h_j$ realizes exactly the AKLT model when $\lambda\rightarrow\infty$ in Eq.~\eqref{H_triplet}.
Note that the mapping from spin-1/2 ladder to spin-1 chain is many-to-one, i.e., there exist other spin-1/2 ladder Hamiltonians that map to AKLT model. In general, however, four-spin terms appear inevitable.

\section{iDMRG simulations of Eq.~\eqref{H_spin1}}
\label{app.Hspin1}
\begin{figure}[h]
    \centering
    \includegraphics[width=0.96\linewidth]{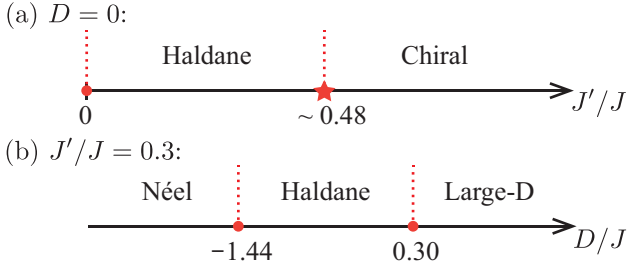}
    \caption{Phase diagrams of $H_{\text{spin-1}}$ in Eq.~\eqref{H_spin1} determined using iDMRG with bond dimension $\chi = 300$ and (a) $D=0$ and (b) $J'/J =0.3$.
    }
    \label{fig:1Dphasediagrams}
\end{figure}

We justify that the Hamiltonian in Eq.~\eqref{H_spin1} realizes a Haldane SPT phase using infinite-size DMRG with bond dimension $\chi=300$. Figure~\ref{fig:1Dphasediagrams}(a) and (b) respectively show our numerically determined phase diagrams at $D=0$ and $J'/J=0.3$.  In \sfigref{fig:1Dphasediagrams}{a}, the $J'=D=0$ point is gapless; turning on $J'>0$ stablizes the Haldane phase, diagnosed by nonzero string order parameters defined in Eq.~\eqref{AKLTstring}.  
For $J'/J\gtrsim 0.48$, the system enters a chiral phase characterized by $\langle \kappa_j \kappa_{j'}\rangle \neq 0$ for $|j-j'|\rightarrow\infty$ where 
\begin{equation}\label{eq:chiralorder}
    \kappa_j = S^x_{j} S^y_{j+1} - S^y_{j} S^x_{j+1}
\end{equation}
defines a chiral order parameter.  
The phase transition near $J'/J\approx0.48$ is studied in Ref.~\onlinecite{kaburagi1999spin}, which reports that when $J'/J>0.490$ there is a gapless chiral phase whereas within the narrow region $0.473<J'/J<0.490$ the system is in a gapped phase with coexisting chiral order and string order. In this work we make use of only the Haldane phase and hence do not include the detailed structure near the transition point. In \sfigref{fig:1Dphasediagrams}{b}, the Haldane phase also has a finite width and resides between the N\'eel phase and the large-$D$ phase---consistent with the phase diagram in Ref.~\onlinecite{tonegawa1995effect} obtained by exact diagonalization.

\section{Numerical results for different $J_z'$}\label{app:numerics}
\begin{figure*}[t]
    \centering
    \includegraphics[width=\linewidth]{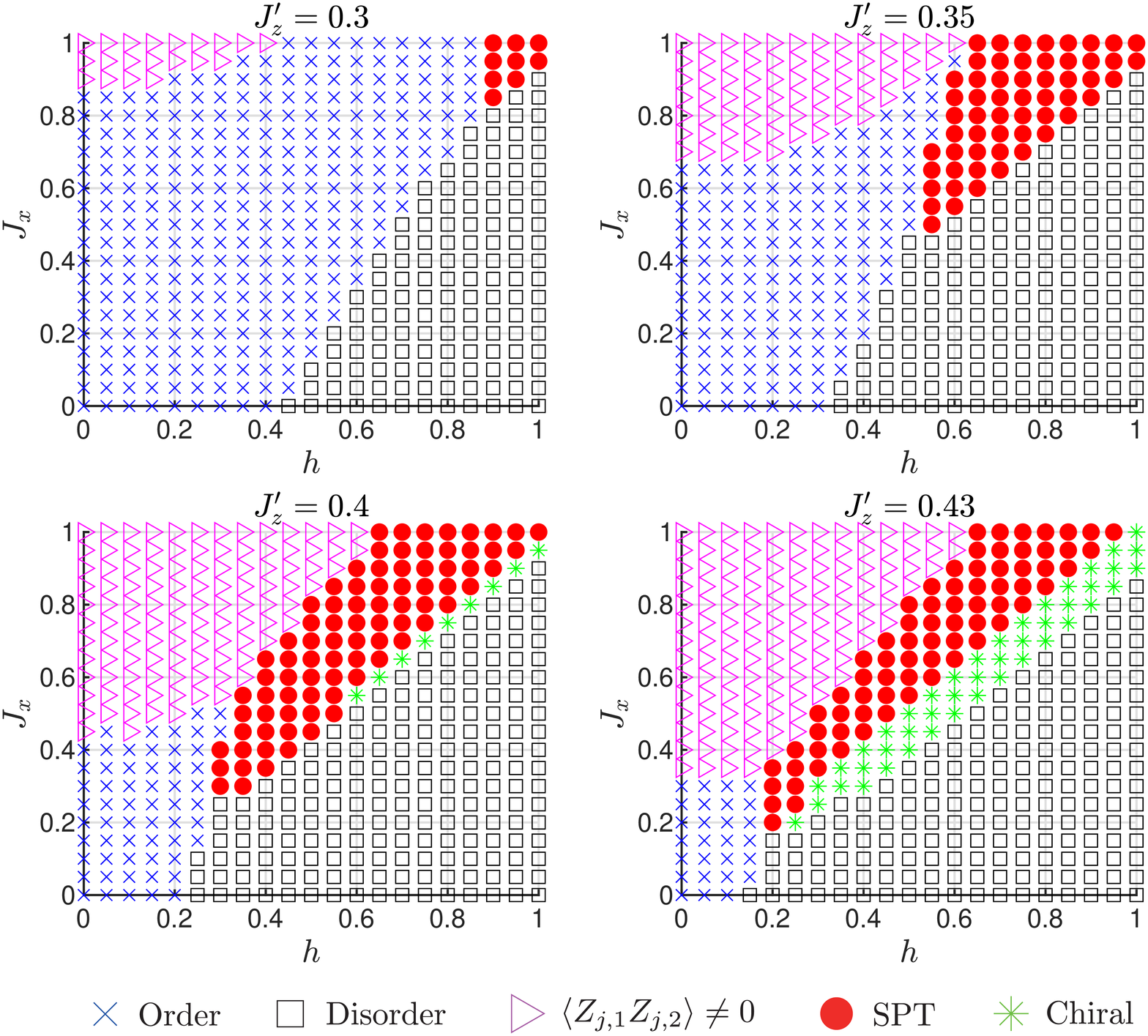}
    \caption{Phase diagrams of Eq.~\eqref{eq:twochain} with $J_z = 1$ for different $J_z'$. The ordered phase is labeled by blue crosses; the disordered phase is labeled by black squares; the partially ordered $\langle Z_{j, 1} Z_{j, 2}\rangle\neq 0$ phase is labeled by pink triangles; the red dots represent the SPT phase; and the green asterisks represent the chiral phase. Note that the range of $J_x$ and $h$ is $[0, 1]$, which differs from Fig.~\ref{fig:twochainPD}. Data were obtained by iDMRG with bond dimension $\chi = 300$. For $J_z'\gtrsim0.45$, DMRG simulations become more difficult to converge and may require larger bond dimension. Since in the main text we focus on the existence of the SPT phase, we leave determination of accurate phase diagrams at $J_z'\gtrsim0.45$ for future work.}
    \label{fig:phasediagramapp}
\end{figure*}

\begin{figure}[h]
    \centering
    \includegraphics[width=\linewidth]{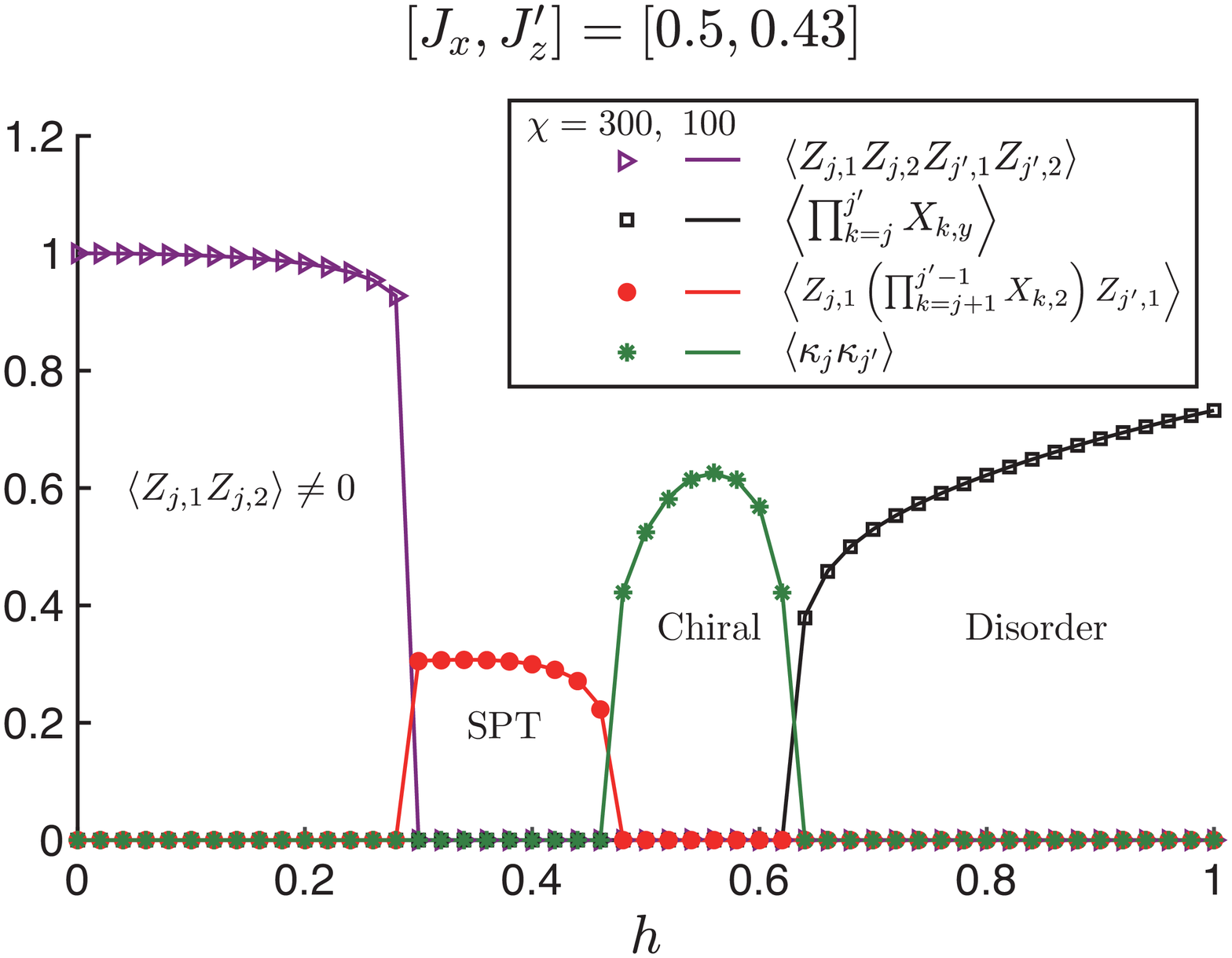}
    \caption{Different order parameters for the two-chain system described by Eq.~\eqref{eq:twochain} with fixed $[J_x, J_z'] = [0.5, 0.43]$ as a function $h$. We evaluate the order parameters using $|j-j'|=600$.  The solid lines were obtained by iDMRG with bond dimension $\chi=100$, while the markers were obtained using $\chi=300$.  Excellent agreement between the data for the two bond dimensions indicates that $\chi=300$ is sufficiently large to identify the phases and locate the phase boundaries to the accuracy presented in the phase diagrams from Fig.~\ref{fig:phasediagramapp}.}
    \label{fig:lineplot}
\end{figure}

Here we investigate the phase diagram of the two-chain model in Eq.~\eqref{eq:twochain} with different $J_z'$ values, thus extending the main-text results from Fig.~\ref{fig:twochainPD}.  We fix $J_z = 1$ throughout this appendix. At the decoupled-chain limit ($J_x = 0$), each chain realizes an axial next-nearest-neighbor Ising (ANNNI) model \cite{Selke1988,Suzuki2013} in a transverse field $h$. When $h=0$, the ground state of the ANNNI model is ordered for $J_z'<0.5$ and has a period of 4 (antiphase) for $J_z'>0.5$. At $J_z'=0.5$, the ground state is exponentially degenerate with degeneracy $\sim \varphi^N$, where $N$ is the system size and $\varphi = (\sqrt{5} + 1)/2 \approx 1.618$ is the golden ratio. The phase diagram of the ANNNI model under a transverse field has been studied with various methods (see, e.g., Chapter 4 of Ref.~\onlinecite{Suzuki2013}). Our goal is to explore the evolution of the phase diagram when the chains couple via $J_x \neq 0$, focusing on the regime $J_z' < 0.5$.

Figure~\ref{fig:phasediagramapp} plots the phase diagrams for $J_z' = \{0.3, 0.35, 0.4, 0.43\}$ obtained using infinite-size DMRG with bond dimension $\chi=300$. Five different phases appear:
\begin{enumerate}
    \item Ordered phase with long range $ZZ$ correlation
    \begin{equation}
        \langle Z_{j, y} Z_{j', y} \rangle \neq 0, \quad y\in \{1,2\},~|j-j'|\rightarrow \infty;
    \end{equation}
    \item Disordered phase with
    \begin{equation}
        \left\langle \prod_{k=j}^{j'} X_{k,y} \right\rangle \neq 0, \quad y\in \{1,2\},~|j-j'|\rightarrow \infty;
    \end{equation}
    \item A partially ordered phase with 
    \begin{equation}
        \langle Z_{j, 1}\rangle =  \langle Z_{j, 2}\rangle = 0, \text{but}~\langle Z_{j, 1} Z_{j, 2}\rangle\neq 0, \quad \forall j;
    \end{equation}
    \item SPT phase with string order parameter
    \begin{equation}
         \mathcal{O}_{1,2}(j, j') \neq 0, \quad |j-j'|\rightarrow \infty,
    \end{equation}
    where $\mathcal{O}_{1,2}$ takes the form in Eqs.~\eqref{eq:twochainsptop1} and \eqref{eq:twochainsptop2};
    \item Gapless chiral phase with
    \begin{equation}
        \left\langle \kappa_j \kappa_{j'}\right\rangle \neq 0, \quad |j-j'|\rightarrow \infty,
    \end{equation}
    where
    \begin{equation}\label{eq:chiraltwochain}
        \kappa_j = Z_{j, 1} Z_{j+1, 2} - Z_{j, 2} Z_{j+1, 1}.
    \end{equation}
\end{enumerate}
A plot of different order parameters along the $[J_x, J_z'] = [0.5, 0.43]$ line is shown in Fig.~\ref{fig:lineplot}, which justifies the existence of different phases and the sufficiency of choosing $\chi=300$.

\begin{figure*}[t]
    \centering
    \includegraphics[width=\linewidth]{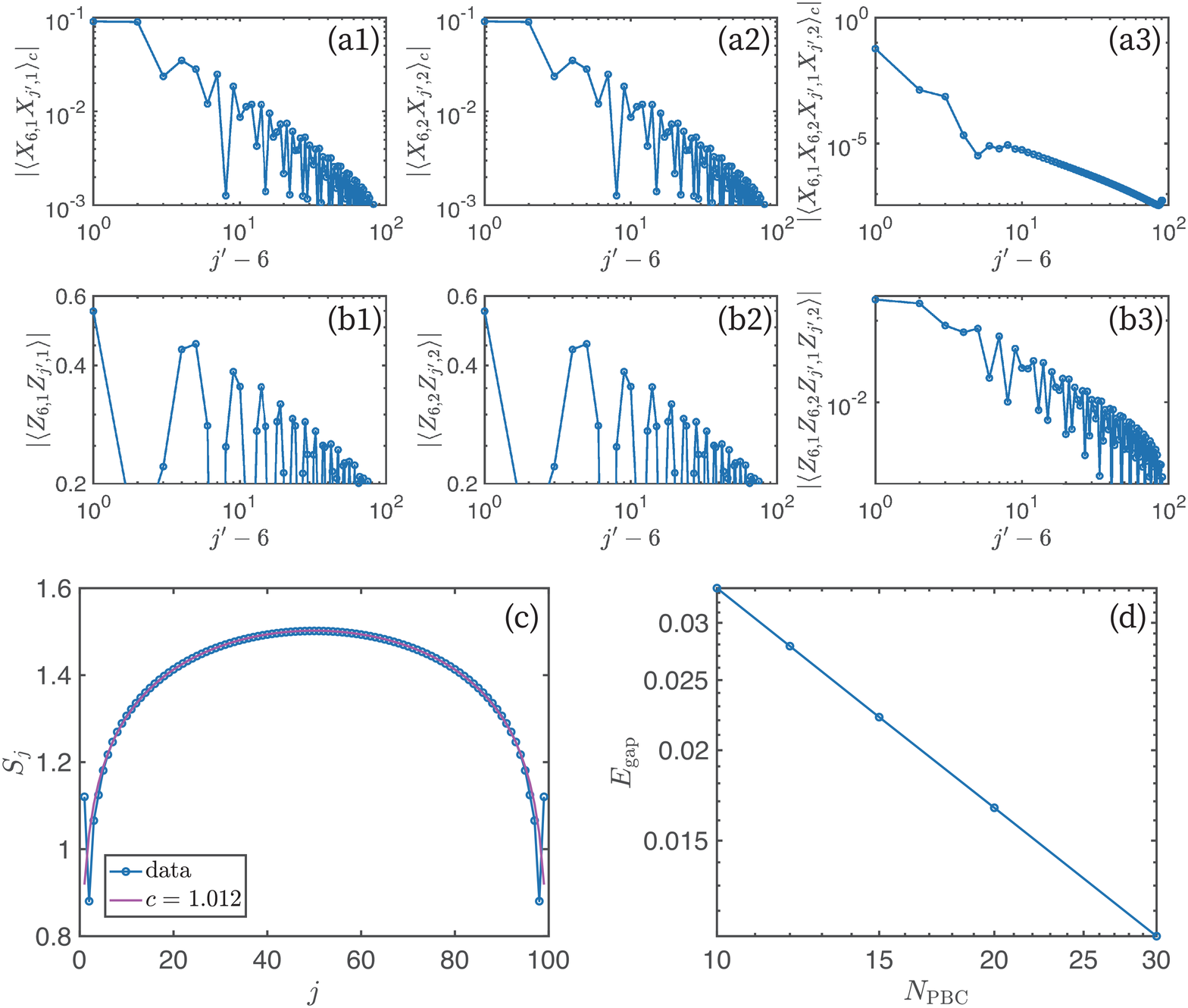}
    \caption{(a-c) DMRG results for an open two-chain system described by Eq.~\eqref{eq:twochain} with $[J_x, h, J_z'] = [0.45, 0.5, 0.43]$, system size $N=100$, and bond dimension $\chi = 300$.  Panels (a1-a3) show various (connected) correlators involving $X$ operators, while (b1-b3) show various $Z$ correlators; all of these data indicate power-law decay.  Panel (c) plots the entanglement entropy $S_j$ as a function of subsystem size $j$. A fit of the formula $S_j = \frac{c}{6} \log \left(\frac{2 N}{\pi} \sin \left(\frac{\pi j}{N}\right)\right)+\text { const }$ implies that the central charge $c$ is 1. (d) DMRG results for the energy gap $E_{\rm gap}$ of a periodic two-chain systems with $[J_x, h, J_z'] = [0.45, 0.5, 0.43]$ as a function of system size $N_{\rm PBC}$, again with $\chi = 300$. We find $E_{\rm gap} \sim 1/ N_{\rm PBC}$, which further indicates gaplessness of the chiral phase. Note that the ground state degeneracy is 2, and that we compute the energy gap above the ground state manifold.}
    \label{fig:chiralphase}
\end{figure*}

All states except for the chiral phase appear already in Fig.~\ref{fig:twochainPD}.  The chiral phase sets in when $J_z'\gtrsim 0.4$; note that the characteristic chiral order defined in Eq.~\eqref{eq:chiraltwochain} reduces to the chiral order for the spin-1 chain [Eq.~\eqref{eq:chiralorder}] under the mapping in Table~\ref{tab.projection}.
The non-zero $\kappa_j$, and vanishing of $\braket{Z_{j,y}}$ and $\braket{Z_{j,1}Z_{j,2}}$, are consistent with a phase that breaks inter-chain reflection symmetry but preserves a $\mathbb Z_4$ subgroup generated by $\prod_j X_{j,1}$ followed by inter-chain reflection. This symmetry is compatible with the twofold degeneracy that we observe (associated with opposite signs for $\kappa_j$).
Moreover, we observe power-law decay among various operators---indicative of a gapless phase---with entanglement entropy revealing a central charge of $c=1$.  Figure~\ref{fig:chiralphase} presents DMRG results (correlation functions, entanglement entropy, and energy gap) for an open two-chain system in the chiral phase with $[J_x, h, J_z'] = [0.45, 0.5, 0.43]$ and system size $N=100$.

Returning to Fig.~\ref{fig:phasediagramapp}, the minimum value of inter-chain coupling $J_x$ at which the SPT phase can form, $\min_{\rm SPT}{J_x}$, clearly varies with $J_z'$.  As $J_z'$ approaches 0.5, the minimum value becomes rather small, e.g., $\min_{\rm SPT}{J_x}\approx 0.2$ for $J_z'=0.43$.  Extrapolating $\min_{\rm SPT}{J_x}$ with $J_z'$ suggests that the SPT may set in at arbitrarily weak $J_x$ at $J_z' \sim 0.5$---though competition from the chiral phase poses a subtlety.  The asymptotic fate of the SPT at $J_z' \rightarrow 0.5$ would be interesting to address in future work.

\section{Kitaev strip} \label{app:kitaevstrip}

\begin{figure}
    \centering
    \includegraphics[width=\linewidth]{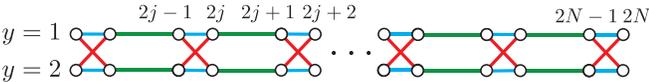}
    \caption{Kitaev strip geometry described by Eq.~\eqref{Hstrip}.  Note the close relationship to the ladder from Fig.~\ref{fig:unfold}(a,b).}
    \label{fig:kitaevstrip}
\end{figure}

Consider spin-1/2 degrees of freedom on the strip geometry shown in Fig.~\ref{fig:kitaevstrip} with Hamiltonian
\begin{equation}
    H_{\rm strip} = \sum_{\text{{blue}}} X_{j,y} X_{j',y'} + \sum_{ \text{{green}}} Y_{j,y}Y_{j',y'} + \sum_{\text{{red}}} Z_{j,y}Z_{j',y'}.
    \label{Hstrip}
\end{equation}
For simplicity we have assigned equal strength for different types of bonds.
This problem closely relates to the Kitaev honeycomb  model 
%, as the latter can be obtained from the former through 
via the unfolding process sketched in Fig.~\ref{fig:unfold}(a,b).  Notice that in Fig.~\ref{fig:kitaevstrip} we assume an even number $2N$ of sites per chain, which facilitates connection to the mean-field description of the SPT discussed in Sec.~\ref{fermionized_sec} and Fig.~\ref{fig:unfold}.  

Like the 2D Kitaev honeycomb model, the strip Hamiltonian enjoys local, mutually commuting conserved quantities. One can check that the following operators (for any $j$) commute with themselves and with the Hamiltonian:
\begin{align}\label{eq:kitaevstrip}
    &Y_{2j-1, 1}Y_{2j, 1}Y_{2j-1, 2}Y_{2j-1, 2}, \nonumber\\
    &Y_{2j-1, 1}Z_{2j, 1}X_{2j+1, 1}Y_{2j+2, 2}Z_{2j+1, 2}X_{2j, 2}.
\end{align}
The 6-spin operator in the second line acts like the plaquette operators in the 2D honeycomb model, which one can directly see from the unfolding process in Fig.~\ref{fig:unfold}. The 4-spin operator in the first line is conserved because the geometry of the strip enables a new type of loop in the honeycomb lattice, e.g., loop $1 \rightarrow 3 \rightarrow 2 \rightarrow 4 \rightarrow 1$ in \sfigref{fig:unfold}{b}.  Additionally, the system preserves two $\mathbb{Z}_2$ symmetries generated by $\prod_{k=1}^{2N} Z_{k,1}$ and $\prod_{k=1}^{2N} Z_{k,2}$, one of which is independent of the local conserved quantities above.

The Kitaev strip can be solved analytically by performing a Jordan-Wigner transformation to Majorana fermion operators
\begin{align}
    & \gamma_{2j-1, y} = \mathcal{G}_y Z_{1, y} Z_{2, y}\cdots Z_{2j-2, y} Y_{2j-1, y}, \nonumber \\
    & \tilde \gamma_{2j-1, y} = \mathcal{G}_y Z_{1, y} Z_{2, y}\cdots Z_{2j-2, y} X_{2j-1, y}, \nonumber \\
    & \gamma_{2j, y} = \mathcal{G}_y Z_{1, y} Z_{2, y}\cdots Z_{2j-1, y} X_{2j, y}, \nonumber \\
    & \tilde \gamma_{2j, y} = \mathcal{G}_y Z_{1, y} Z_{2, y}\cdots Z_{2j-1, y} Y_{2j, y},
\end{align}
where $\mathcal{G}_{y=1} = 1$ and $\mathcal{G}_{y=2} = \prod_{k=1}^{2N} Z_{k, 1}$ ensure proper anticommutation relations. In this representation $-i \gamma_{2j-1, y} \gamma_{2j, y}$ acts along the $x$-links (blue links in Fig.~\ref{fig:kitaevstrip}); $i \gamma_{2j, y} \gamma_{2j+1, y}$ acts along the $y$-links (green links in Fig.~\ref{fig:kitaevstrip}); and $\gamma_{2j-1, 1} \tilde \gamma_{2j-1, 1} \gamma_{2j, 2} \tilde \gamma_{2j, 2}$ and $\gamma_{2j, 1} \tilde \gamma_{2j, 1} \gamma_{2j-1, 2} \tilde \gamma_{2j-1, 2}$ act along the $z$-links (red links in Fig.~\ref{fig:kitaevstrip}). The terms $i \tilde \gamma_{2j-1, 1} \tilde \gamma_{2j, 2}$ and $i \tilde \gamma_{2j-1, 2} \tilde \gamma_{2j, 1}$ commute with the Hamiltonian and define a $\mathbb{Z}_2$ gauge field. The mean-field description of the two-chain model in Sec.~\ref{fermionized_sec} corresponds to the $\pi$-flux sector in this gauge theory,  wherein $Y_{2j-1, 1}Y_{2j, 1}Y_{2j-1, 2}Y_{2j-1, 2} = -1$ for all $j$.

For an open system with length $2N$, we find using exact diagonalization that the ground states are $2^N$-fold degenerate and satisfy $Y_{2j-1, 1}Y_{2j, 1}Y_{2j-1, 2}Y_{2j-1, 2} = +1$ for all $j$. (One can understand the degeneracy by observing that for each $j$, either $i\tilde{\gamma}_{2j-1,1}\tilde{\gamma}_{2j,2}$ or $i\tilde{\gamma}_{2j-1,2}\tilde{\gamma}_{2j,1}$ equals $-1$; the minus signs can be tiled in exponentially many ways.) 
 Therefore the $\pi$-flux sector is not accessible in the ground state manifold of the pure model in Eq.~\eqref{Hstrip}.  However, the $\pi$-flux sector can be stabilized by adding a term $\delta \sum_{j} Y_{2j-1, 1}Y_{2j, 1}Y_{2j-1, 2}Y_{2j-1, 2}$ to Eq.~\eqref{Hstrip} with $\delta>0$ exceeding some threshold. Upon including such a term, we find a fourfold ground-state degeneracy with the ground states satisfying $i \tilde \gamma_{2j-1, 1} \tilde \gamma_{2j, 2} = i \tilde \gamma_{2j-1, 2} \tilde \gamma_{2j, 1}$ and $Y_{2j-1, 1}Y_{2j, 1}Y_{2j-1, 2}Y_{2j-1, 2} = -1$. One can also check that with periodic boundary conditions the ground state becomes unique, suggesting that the degeneracy with open boundary conditions comes from edge modes. Given the connection to the mean-field description of the two-chain SPT in Sec.~\ref{fermionized_sec}, we conclude that the ground state of this Kitaev strip perturbed by the $\delta$ term above realizes a $\mathbb{Z}_2 \times \mathbb{Z}_2$ SPT. 

An alternative type of Kitaev ladder system shown in Fig.~\ref{fig:unfold}(c) emerges from imposing periodic boundary conditions on the honeycomb lattice in a distinct way, and has been discussed, e.g., in Refs.~\onlinecite{Catuneanu19,DeGottardi_2011}. The system  forms a string of squares that lead to loop operators of the form $X_{j, 1} X_{j+1, 1} Y_{j, 2} Y_{j+1, 2}$ and $Y_{j, 1} Y_{j+1, 1} X_{j, 2} X_{j+1, 2}$ that commute with the Hamiltonian. 
This type of Kitaev ladder, when similarly described by Eq.~\eqref{Hstrip} but with the new pattern of colored bonds, does not host an SPT. Instead, our DMRG simulations capture a symmetry-breaking phase with order parameter $X_{j,1} Y_{j,2} - X_{j,2} Y_{j,1} \neq 0$.  We indeed find twofold ground state degeneracy for both open and periodic boundary conditions, indicating that this kind of Kitaev ladder does not support nontrivial edge modes.

\bibliographystyle{apsrev4-2}
\bibliography{ref}
\end{document}